\documentclass[aps,prb,reprint,groupedaddress]{revtex4-1}
\usepackage{graphicx}
\usepackage{subfigure}
\begin{document}


\title{Lattice density-functional theory on graphene}


\author{Mari Ij\"{a}s}
\email[]{mari.ijas@tkk.fi}
\author{Ari Harju}
\affiliation{Department of Applied Physics and Helsinki Institute of Physics, Aalto University, FI-02150 Espoo, Finland}


\date{\today}

\begin{abstract}
 A density-functional approach on the hexagonal graphene lattice is developed using an exact numerical solution to the Hubbard model as the reference system. Both nearest-neighbour and up to third nearest-neighbour hoppings are considered and exchange-correlation potentials within the local density approximation are parameterized for both variants. The method is used to calculate the ground-state energy and density of graphene flakes and infinite graphene sheet. The results are found to agree with exact diagonalization for small systems, also if local impurities are present. In addition, correct ground-state spin is found in the case of large triangular and bowtie flakes out of the scope of exact diagonalization methods.
\end{abstract}


\maketitle

\section{Introduction}

Density-functional theory (DFT) is one of the most widely-used tools in the field of electronic structure simulations. Based on the Hohenberg-Kohn theorems \cite{Hohenberg-Kohn} and Kohn-Sham equations,\cite{Kohn-Sham} it allows one to treat interacting many-body problems as a collection of one-body problems with an effective potential. The many-body effects are taken into account through this exchange-correlation potential, whose functional form is not known exactly. 

Despite its apparent simplicity, the Hubbard model, containing only the on-site contribution of the Coulomb interaction, has been found to describe a wide range of correlated materials and phenomena.\cite{Tasaki} As the model cannot be solved analytically above one dimension, numerical or approximative methods, such as exact diagonalization (ED) on small systems, quantum Monte Carlo or perturbation theory, have to be applied. For example in transport calculations on graphene, the mean-field approximation has succesfully been applied to allow treatment of systems consisting of few hundreds of atoms.\cite{Hancock} 

The DFT and the Hubbard model can be combined in several ways.\cite{Capelle} The most obvious choices are either to determine the model parameters based on first principles calculations or to incorporate a Hubbard-type interaction into the DFT exchange-correlation functional, resulting in the so-called DFT+U method.\cite{Anisimov} The model, however, can also be considered as an interesting system on its own and studied using DFT in a lattice formulation. This approach has previously been chosen by, for instance, Capelle \emph{et al.} (named BA-LDA),\cite{Capelle, Lima-Silva} and Gunnarsson and Sch\"{o}nhammer (named SOFT) \cite{Schonhammer-Gunnarsson-Noack, Schonhammer-Gunnarsson, Gunnarsson-Schonhammer} and Schindlmayr and Godby.\cite{Schindlmayr-Godby} 

The exchange-correlation (XC) functional, the core of the many-body treatment in DFT, is naturally needed also in the lattice formulation. In the case of the one-dimensional Hubbard model, an exact analytical solution based on the Bethe Ansatz is known \cite{Lieb-Wu} and it can be used to parameterize the XC functional.\cite{Lima-Silva} One-dimensional Hubbard chains have been widely studied within this framework.\cite{Lima-Silva, Capelle-Oliveira, Schonhammer-Gunnarsson-Noack, Schonhammer-Gunnarsson, Gunnarsson-Schonhammer, Schindlmayr-Godby} The method has been found to accurately reproduce the exact ground-state energy, also in the presence of impurities modelled as on-site energies.\cite{Lima-Silva}

Also a related method, density matrix functional theory, has been applied on the Hubbard model, not only for the 1-dimensional chain but also for small clusters and square lattice in two and three dimensions.\cite{Lopez-Sandoval, Lopez-Pastor, Lopez-Pastor2}  The functional in this approach has been based on quantum Monte Carlo reference data. In general, the density matrix functional theory has shown a great potential, being applicable even in strongly correlated quantum Hall droplets.\cite{Tolo} 

To the best of our knowledge, no lattice density-functional theory (LDFT) studies have previously been performed on the two-dimensional hexagonal lattice, or in dimensions above one, in general. The LDFT method could be used in transport calculations on graphene instead of the usual mean-field treatment, thus improving the description of correlations for large graphene systems. It would also enable one to study dynamic phenomena in large systems in the form of time-dependent LDFT.  

Graphene nanoflakes, or graphene quantum dots have been proposed as elements of future nanoelectronics devices. Especially triangular nanoflakes have been a subject of recent research interest. Zigzag-edged flakes have been found to show non-trivial spin order and they have been proposed to function for instance as logic devices in nanoelectronics.\cite{Wang-Yazuev, Wang-Meng, Yazuev-Wang} Also the effect of edge termination on their transport properties has recently been studied, spin valve and spin rectification properties being reported.\cite{Sahin} The transmission through triangular junctions was found tunable through holes in the triangular flakes forming the junction.\cite{Chen} They do exhibit spin-polarized ground-states, both in density-functional theory calculations \cite{Wang-Yazuev, Wang-Meng, Yazuev-Wang} and in theoretical considerations within the nearest-neighbour tight-binding scheme in the absence of interactions.\cite{ Fajtlowicz, Potasz}  Our purpose is to apply the Hubbard-based LDFT method on large trianglar nanoflakes, also including the up to third nearest-neighbour hopping neglected in the earlier calculations,\cite{Potasz, Manninen, Akola, Heiskanen} and show that our method captures the essentials of these phenomena. 

\section{Lattice density-functional theory \label{sec:ldft}}

In DFT, density is the main variable instead of the wave function. In the lattice formulation, the discrete density $\{n_i\}$, also called the site occupations, takes this place. The Hohenberg-Kohn theorems apply also in the discrete formulation \cite{Schindlmayr-Godby} and the Kohn-Sham equations read
\begin{equation} \label{eq:KS} \hat{H}_{KS} \psi_i = \epsilon_i \psi_i, \end{equation}
where $\hat{H}_{KS}$, the Kohn-Sham Hamiltonian, is 
\begin{equation} \label{eq:KSH} \hat{H}_{KS} = \hat{T}+\hat{V}_{\mathrm{ext}}+ \hat{V}_H + \hat{V}_{\mathrm{xc}}. \end{equation}
In these equations, $\psi_i$ and $\epsilon_i$ are the Kohn-Sham eigenstates and -energies, respectively, $\hat{T}$ is the kinetic energy operator, $\hat{V}_{\mathrm{ext}}$ the external potential, $\hat{V}_H$ the Hartree potential and $\hat{V}_{\mathrm{xc}}$ the exchange-correlation potential. These are simply defined on the set of lattice sites, instead of as a continuous function of the position variable, in the discrete formulation. As the exchange-correlation potential is density-dependent, $\hat{V}_{\mathrm{xc}} (\{n_i\})$, Eqs.~(\ref{eq:KS}) have to be solved self-consistently. Inclusion of spin in the model leads to separate Kohn-Sham equations for each spin species that are coupled through $\hat{V}_{\mathrm{xc}}$.

The Hubbard model is given by
\begin{equation} \label{eq:Hub} \hat{H} = \sum_{\langle i,j \rangle, \sigma}  (t_{ij}\hat{c}_{i\sigma}^{\dagger}\hat{c}_{j\sigma} + t_{ij}^{\dagger}\hat{c}_{j\sigma}^{\dagger}\hat{c}_{i\sigma}) + U \sum_i \hat{n}_{i\uparrow}\hat{n}_{i\downarrow}, \end{equation}
where $\hat{c}_{i\sigma}$ ($\hat{c}_{i\sigma}^{\dagger}$) are the usual annihilation (creation) operators for spin $\sigma$, $\hat{n}_{i\sigma}$ is $\hat{c}_{i\sigma}^{\dagger}\hat{c}_{i\sigma}$, $t_{ij}$ is the hopping amplitude between sites $i$ and $j$ and $U$ is the strength of the on-site interaction between opposite spins. The set of hopping amplitudes $\{t_{ij}\}$ determines the geometry of the system. In the case of graphene, hoppings up to the first or third neighbour on the hexagonal lattice are routinely included, given by either $t_1 = -2.7$ eV in the case of nearest-neighbour hopping or $t_1 = -2.7$ eV, $t_2 = -0.2$ eV and $t_3 = -0.18$ eV in the case of up to third-nearest neighbour hopping.\cite{Hancock, Hancock-Uppstu} In this work, we choose our unit of energy to be $t =-t_1$, scaling the other hopping amplitudes accordingly. 

The kinetic term $\hat{T}$ of Eq. (\ref{eq:KSH}) is given in our lattice formulation by the hopping term of the Hubbard Hamiltonian. The Hartree potential $\hat{V}_H$, of which only the on-site contribution is taken into account in the Hubbard model, is considered in the sense of the mean-field Hartree-Fock approximation (HF), and the corresponding Hamiltonian is
\begin{equation} \label{eq:HF} \hat{H}_{HF} = \sum_{\langle i,j \rangle, \sigma} t_{ij}(\hat{c}_{i\sigma}^{\dagger}\hat{c}_{j\sigma} + \hat{c}_{j\sigma}^{\dagger}\hat{c}_{i\sigma}) + U\sum_{i\sigma}\hat{n}_{i\sigma}n_{i-\sigma}, \end{equation}
where $n_{i\sigma}$ the electron density of electrons with spin $\sigma$ on the site $i$. The second term in Eq. (\ref{eq:HF}) gives the interaction contribution.

We assume $\hat{V}_{\mathrm{xc}}$ to depend only on the total density $n_i = n_{i\uparrow} + n_{i\downarrow}$ and not separately on the spin-resolved densities. In the LDFT calculation, the Kohn-Sham orbitals $\psi_{i\sigma}$ are solved from Eq. (\ref{eq:KS}) and the new density is calculated from the occupied orbitals of each spin species, 
\begin{equation} \label{eq:den} n_{i\sigma} =\sum_{j\sigma \in occ} |\psi_{i\sigma}^{j\sigma}|^2. \end{equation} 
The new XC potential is then determined based on the occupations and the system is iterated until convergence of the site occupations $\{n_{i\sigma}\}$. If necessary, the convergence may be facilitated using a standard mixing procedure in which the new density is calculated as a linear combination of the current and old density. After convergence, the total energy of the ground-state is obtained from the Kohn-Sham eigenenergies and density as \cite{Thijssen}
\begin{eqnarray} \label{eq:E0} E &=& \sum_{j\in occ, \sigma} \epsilon_{i\sigma} + E_{\mathrm{xc}}(\{n_i\}, U) \nonumber \\ & & -\sum_i n_{i}\hat{V}_{\mathrm{xc}}(n_i,U)-U\sum_i n_{i\uparrow}n_{i\downarrow}. \end{eqnarray}

\begin{figure}
 \subfigure[]{\includegraphics[scale=0.25]{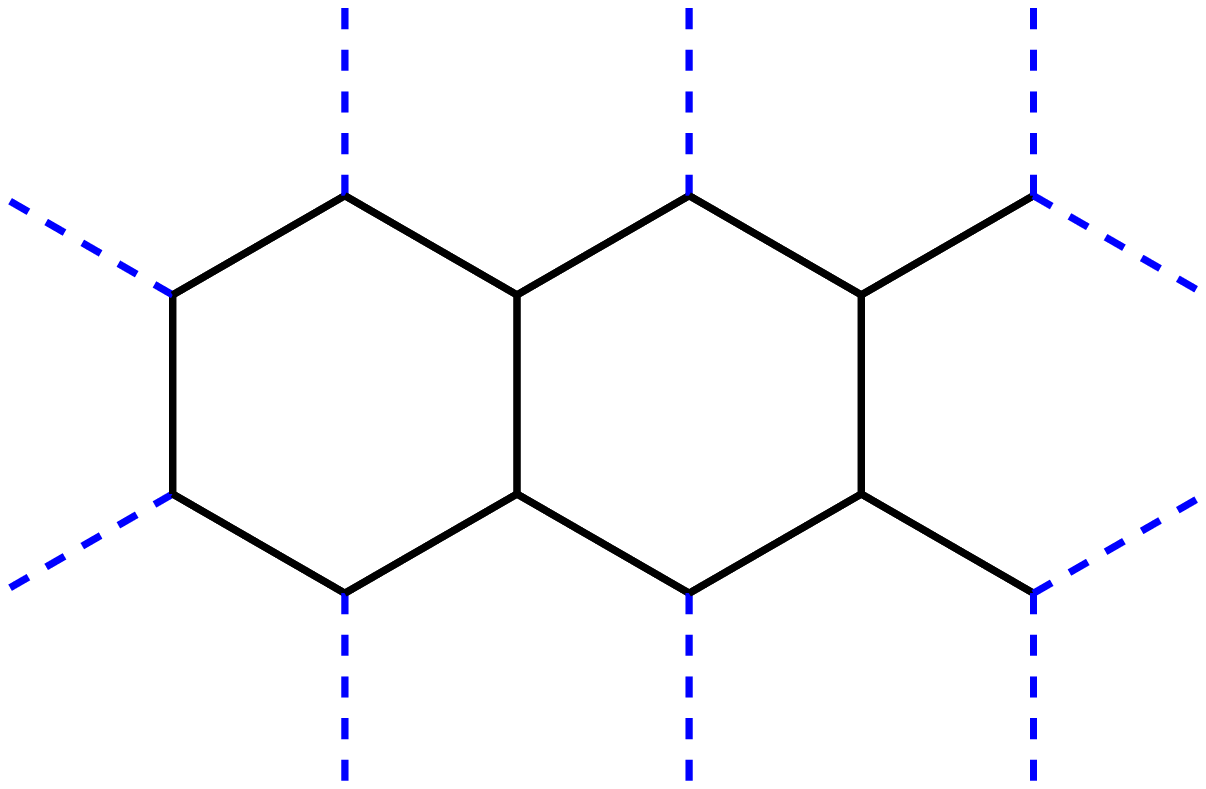}}
 \subfigure[]{\includegraphics[scale=0.25]{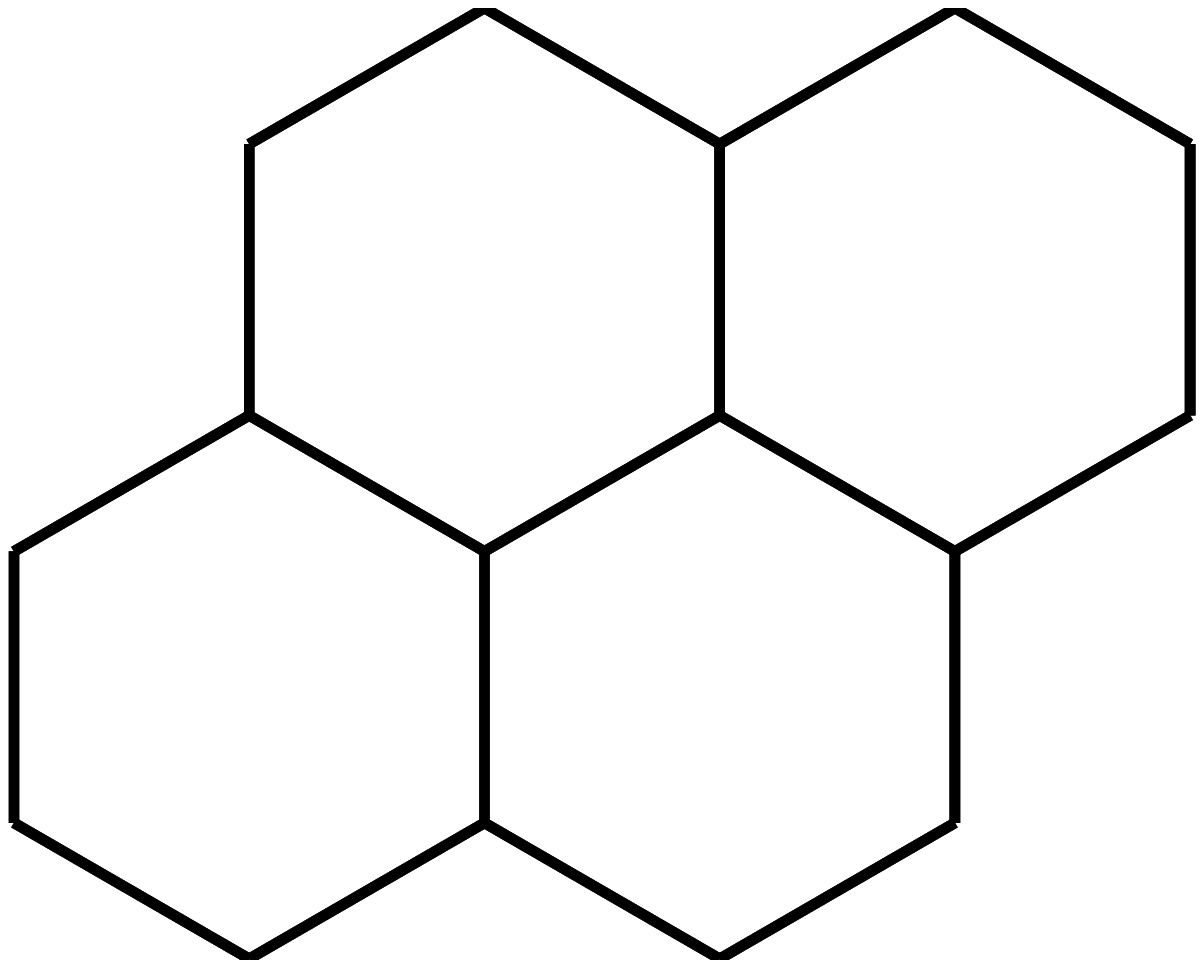}}
\caption{\label{fig:refsys} (a) The 12-atom supercell of the infinite graphene sheet, chosen for the determination of the exchange-correlation energy and potential. The dashed lines show the connections over the periodic boundaries. (b) A C$_{16}$ flake chosen as the first test system of our LDFT method. }
\end{figure}

\section{Determining the XC functional \label{sec:exc}}

\begin{figure*}
		\subfigure[]{\includegraphics[scale=0.25]{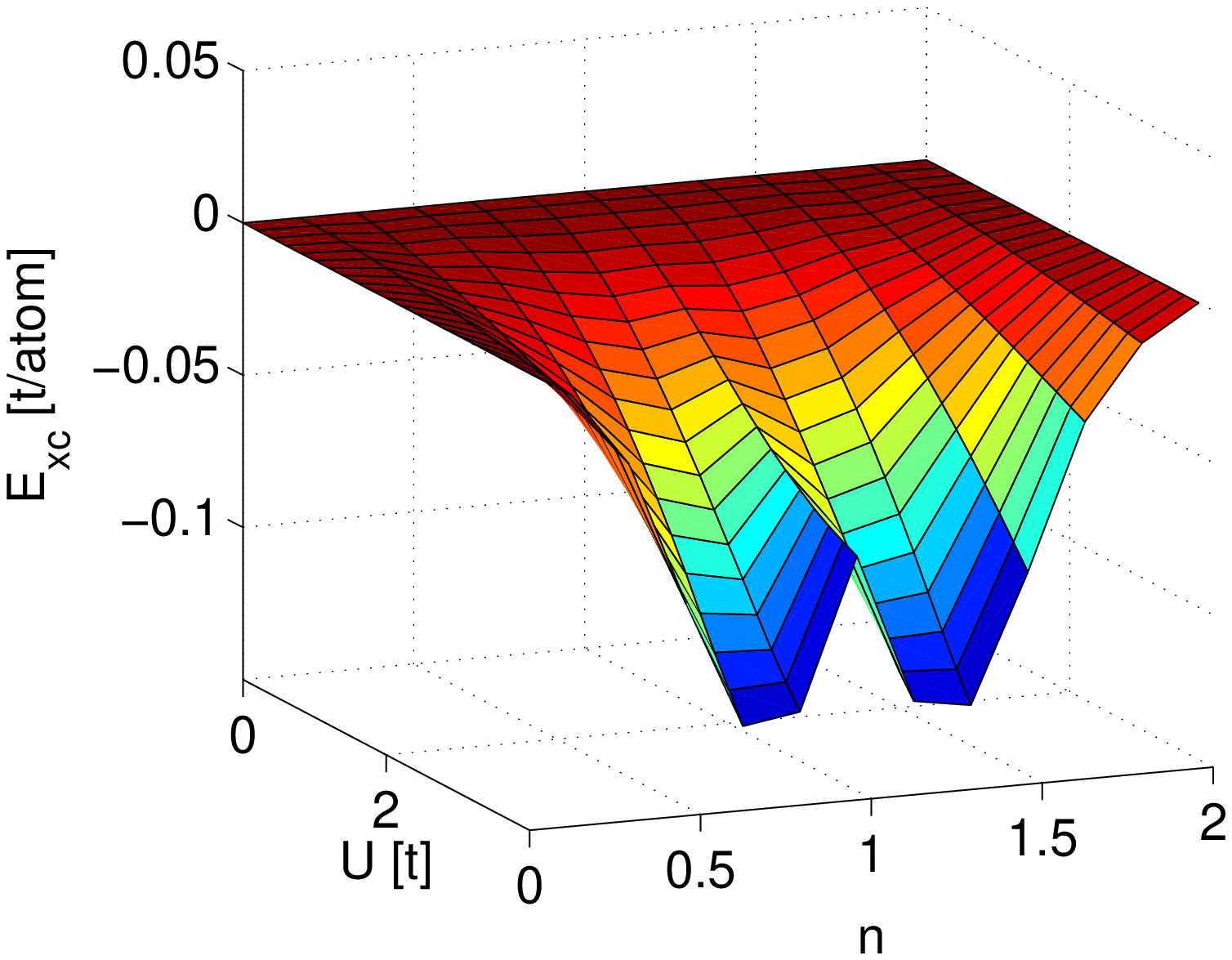}}
		\subfigure[]{\includegraphics[scale=0.25]{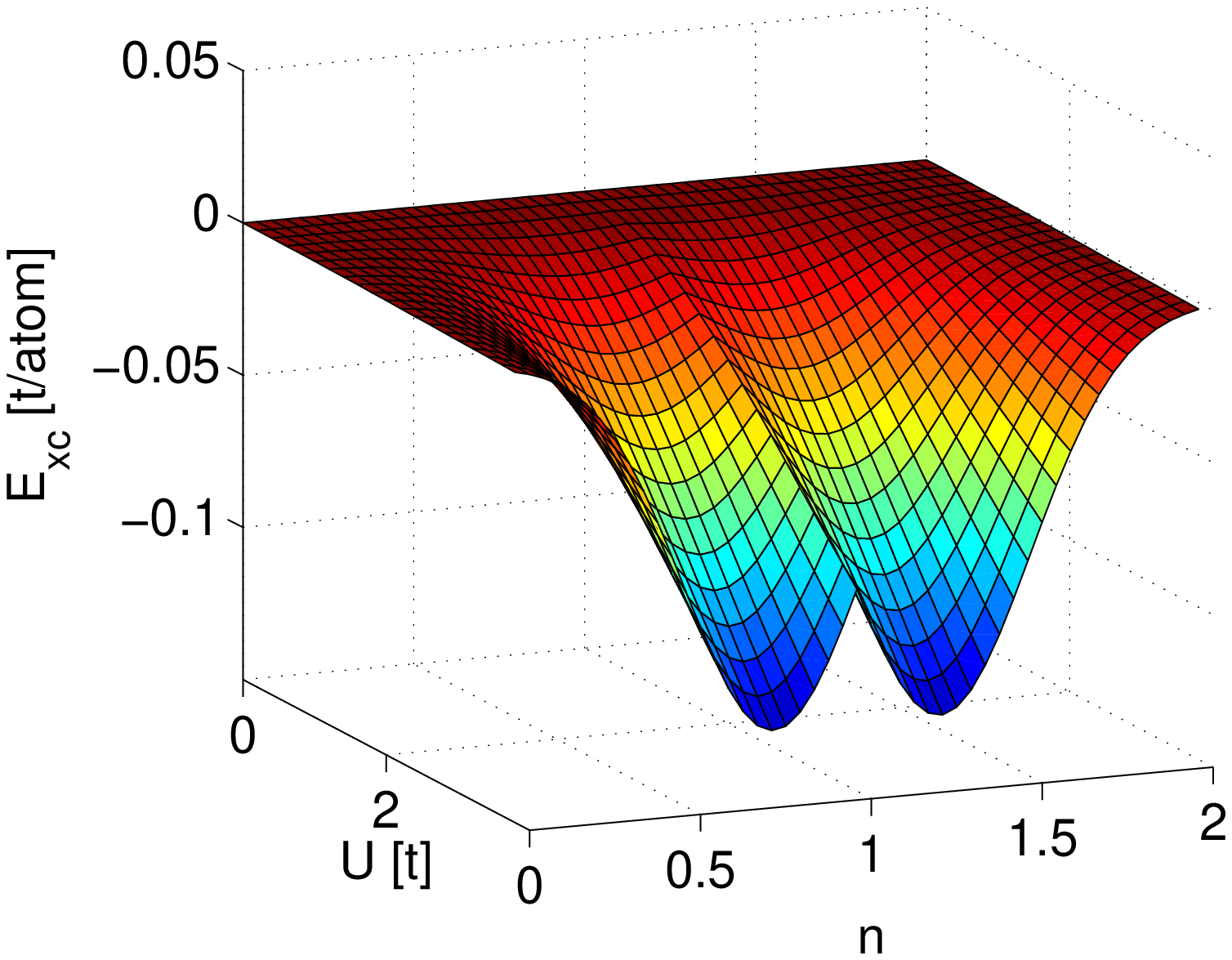}}
		\subfigure[]{\includegraphics[scale=0.25]{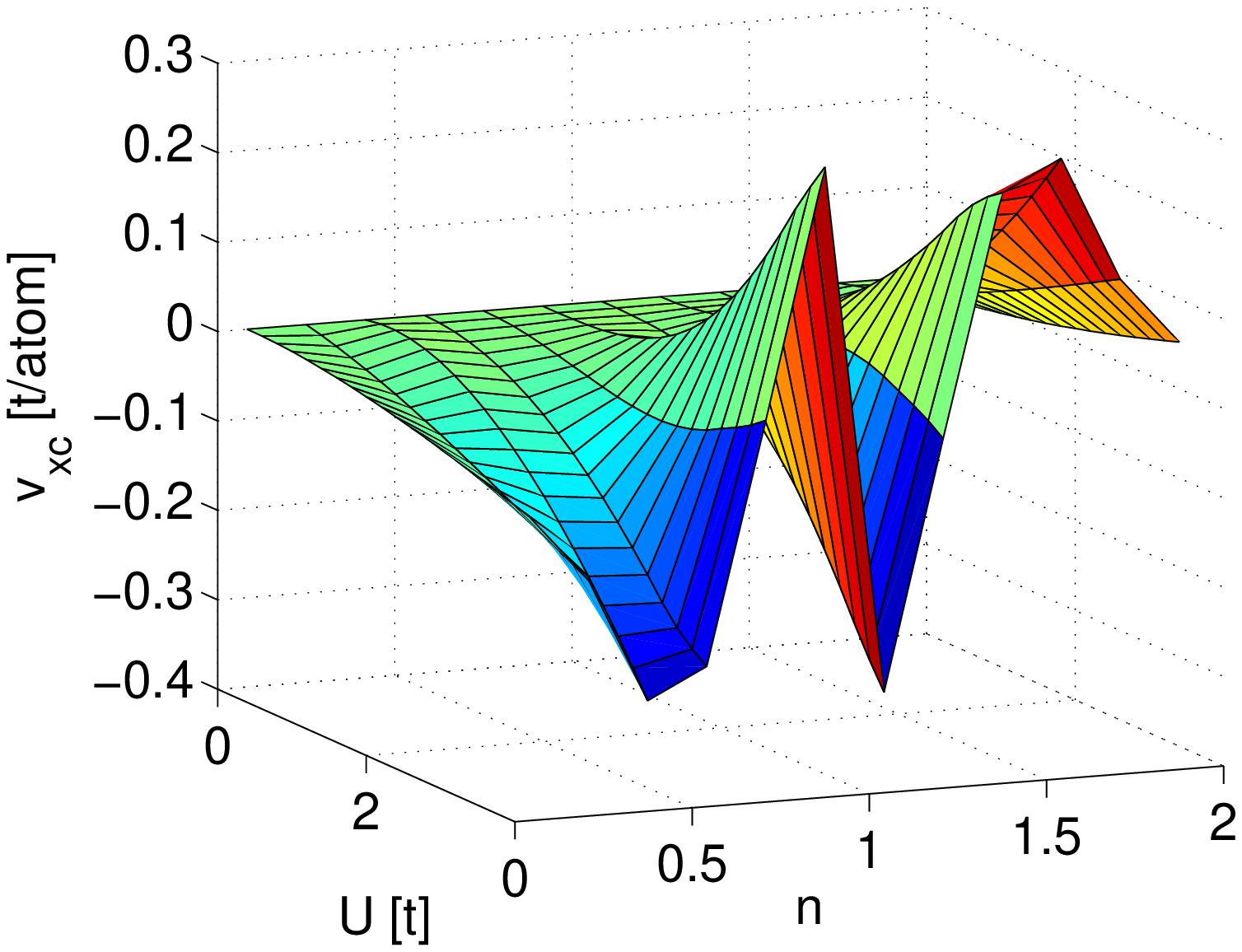}}
		\subfigure[]{\includegraphics[scale=0.25]{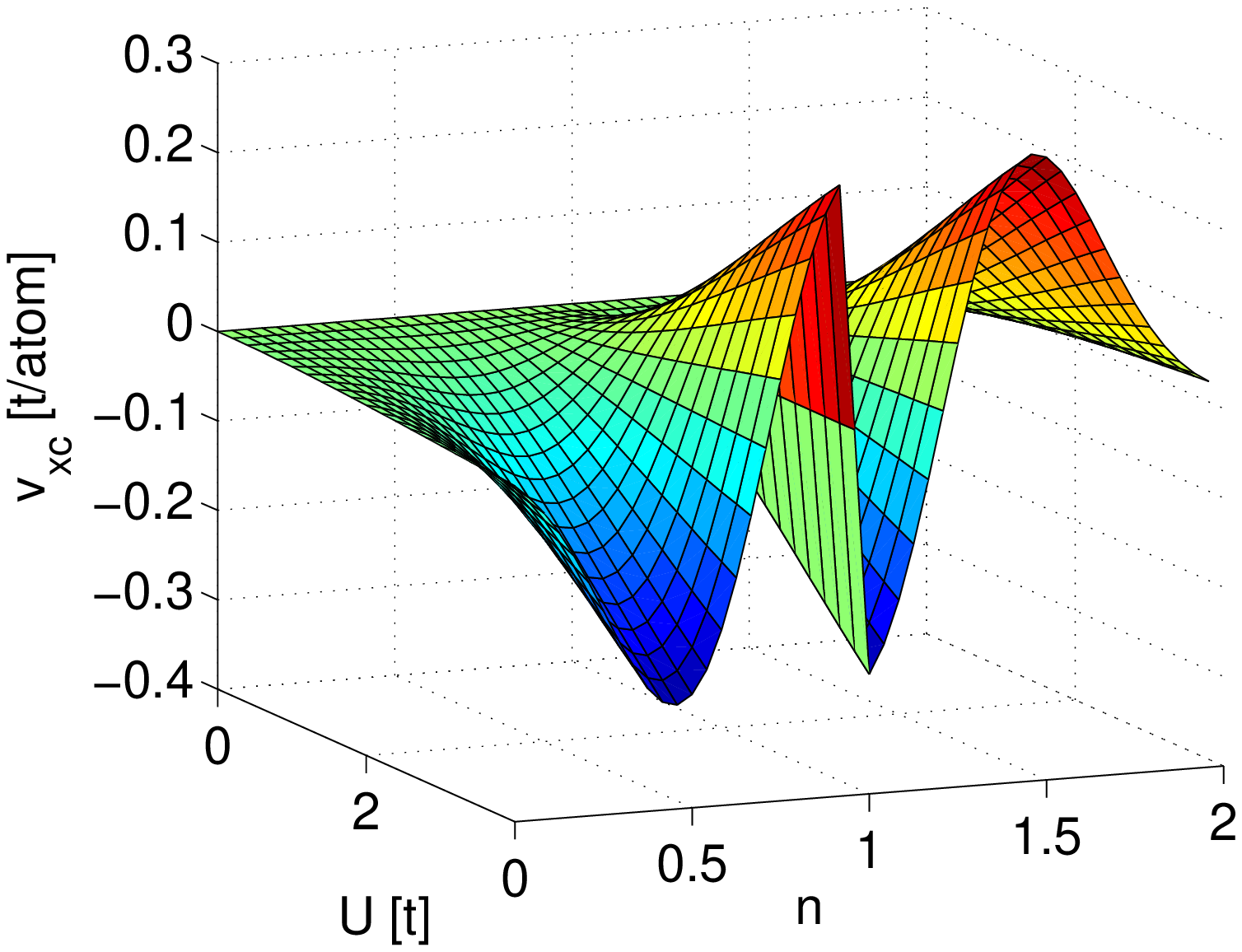}} \\
		\subfigure[]{\includegraphics[scale=0.25]{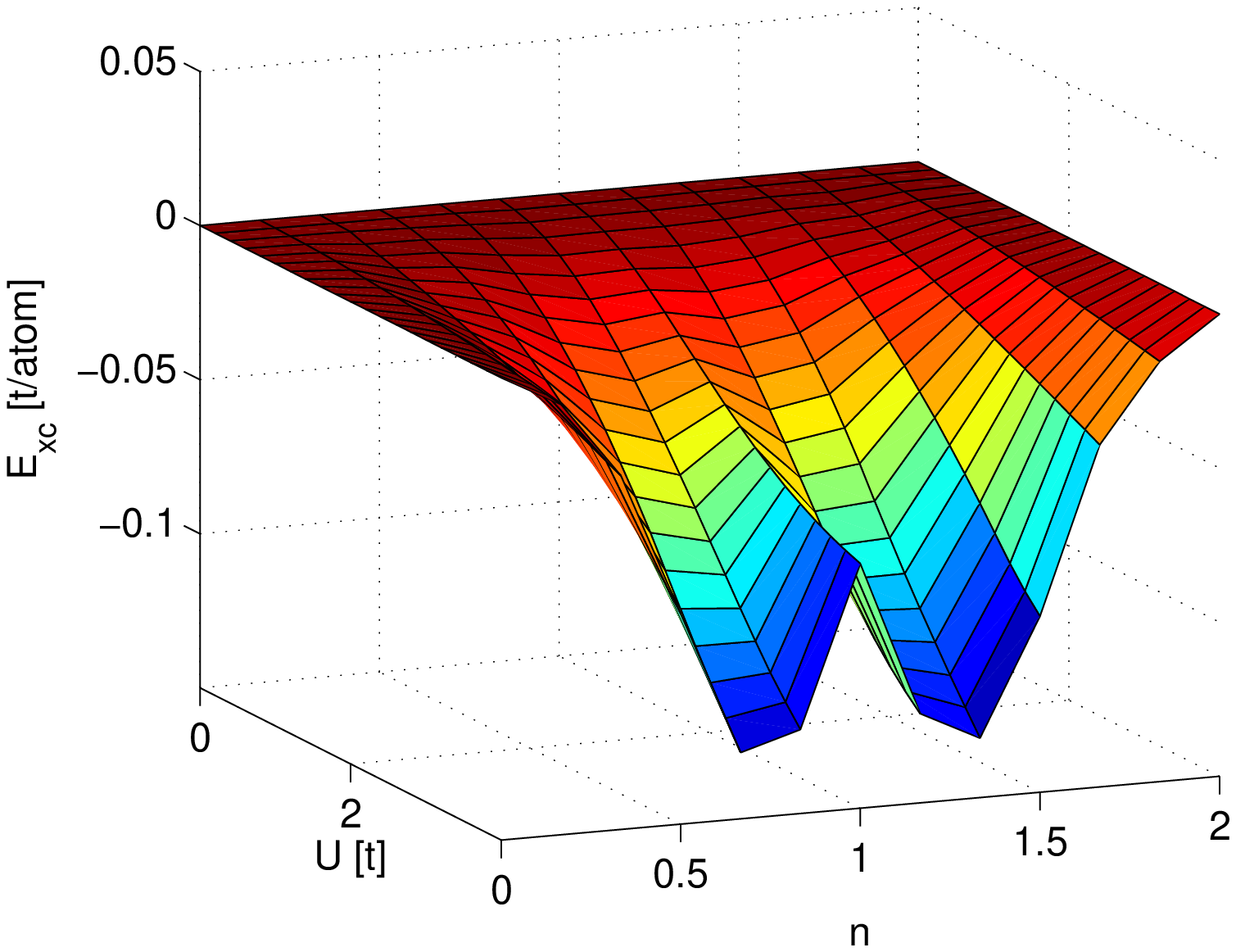}}
		\subfigure[]{\includegraphics[scale=0.25]{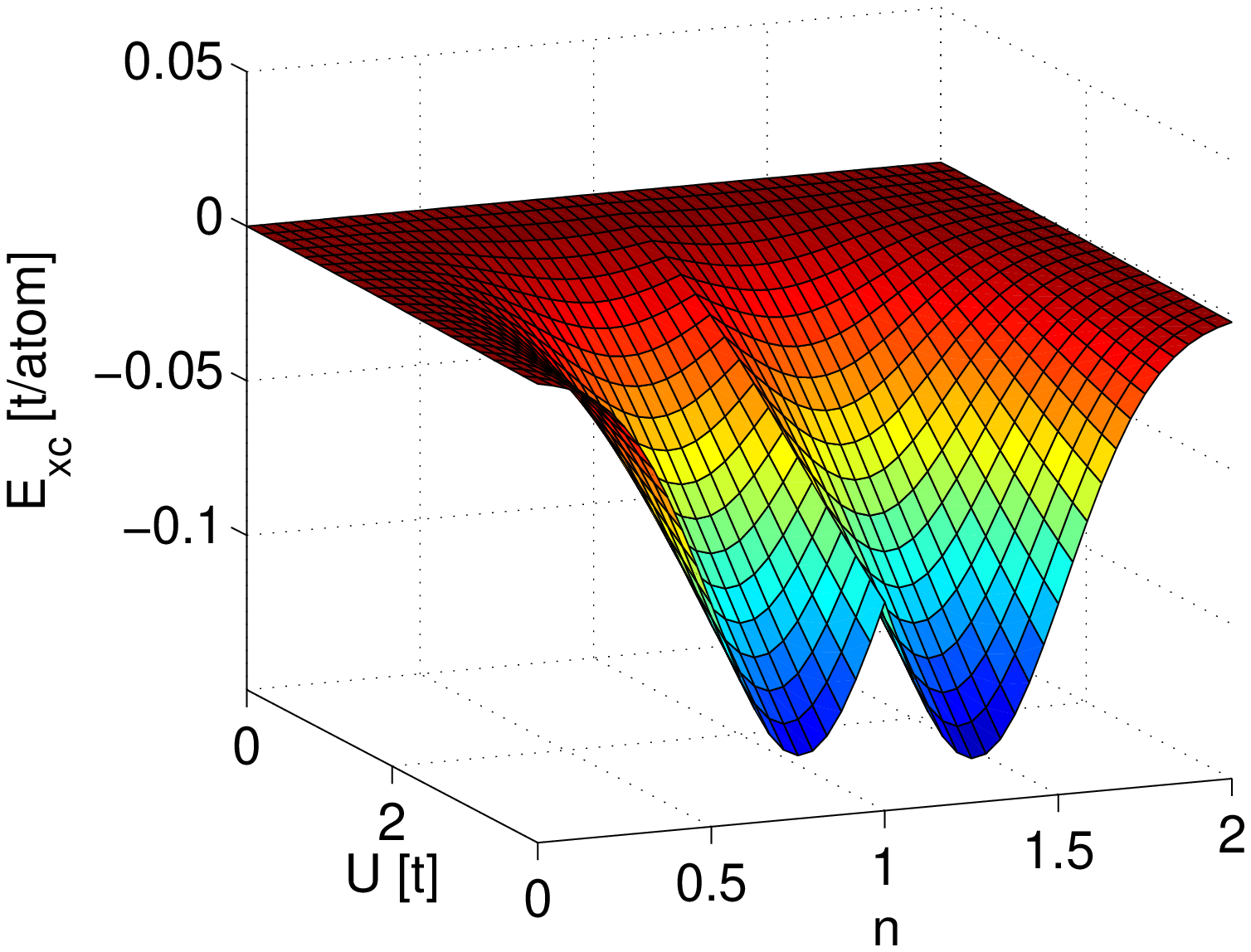}}
		\subfigure[]{\includegraphics[scale=0.25]{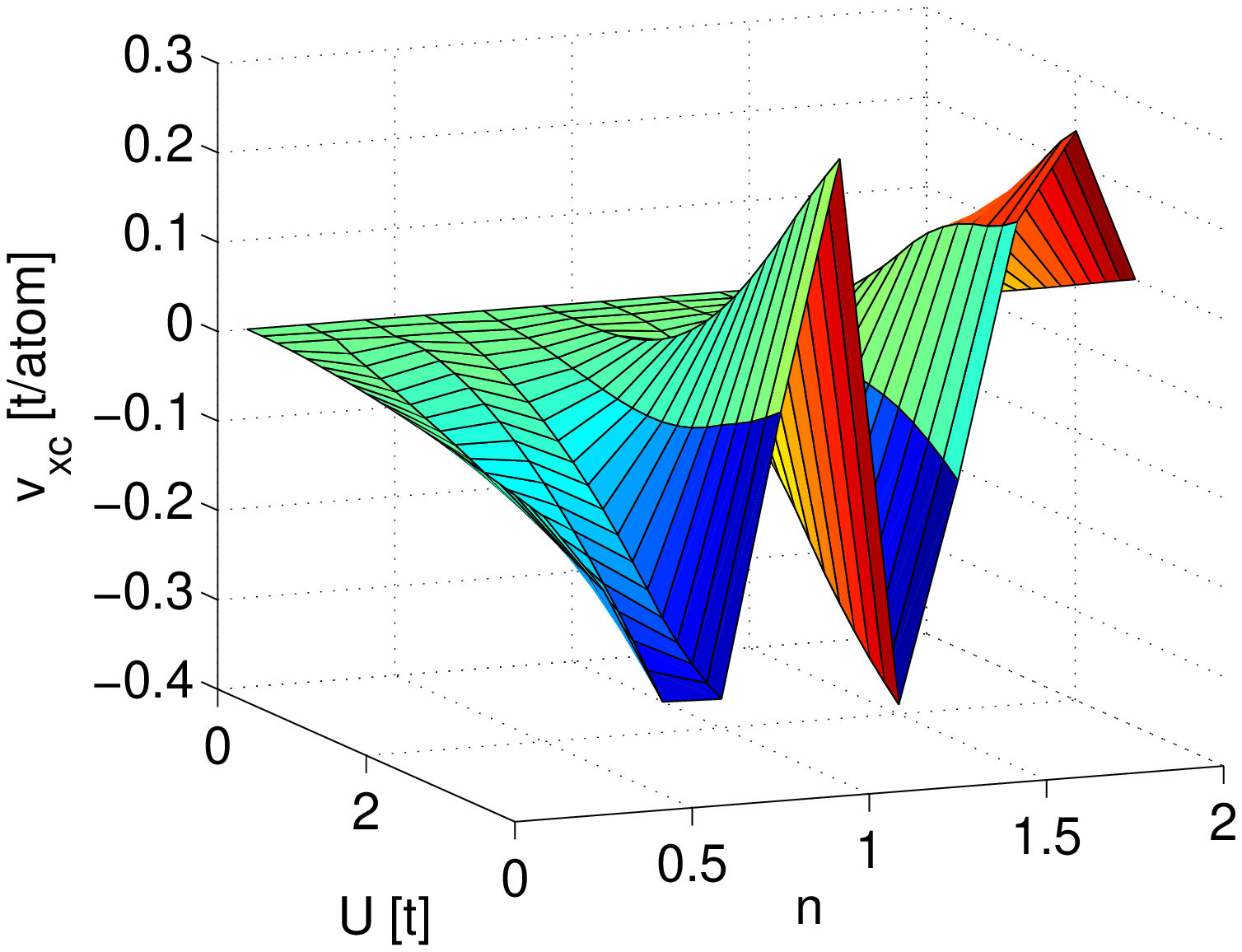}}
		\subfigure[]{\includegraphics[scale=0.25]{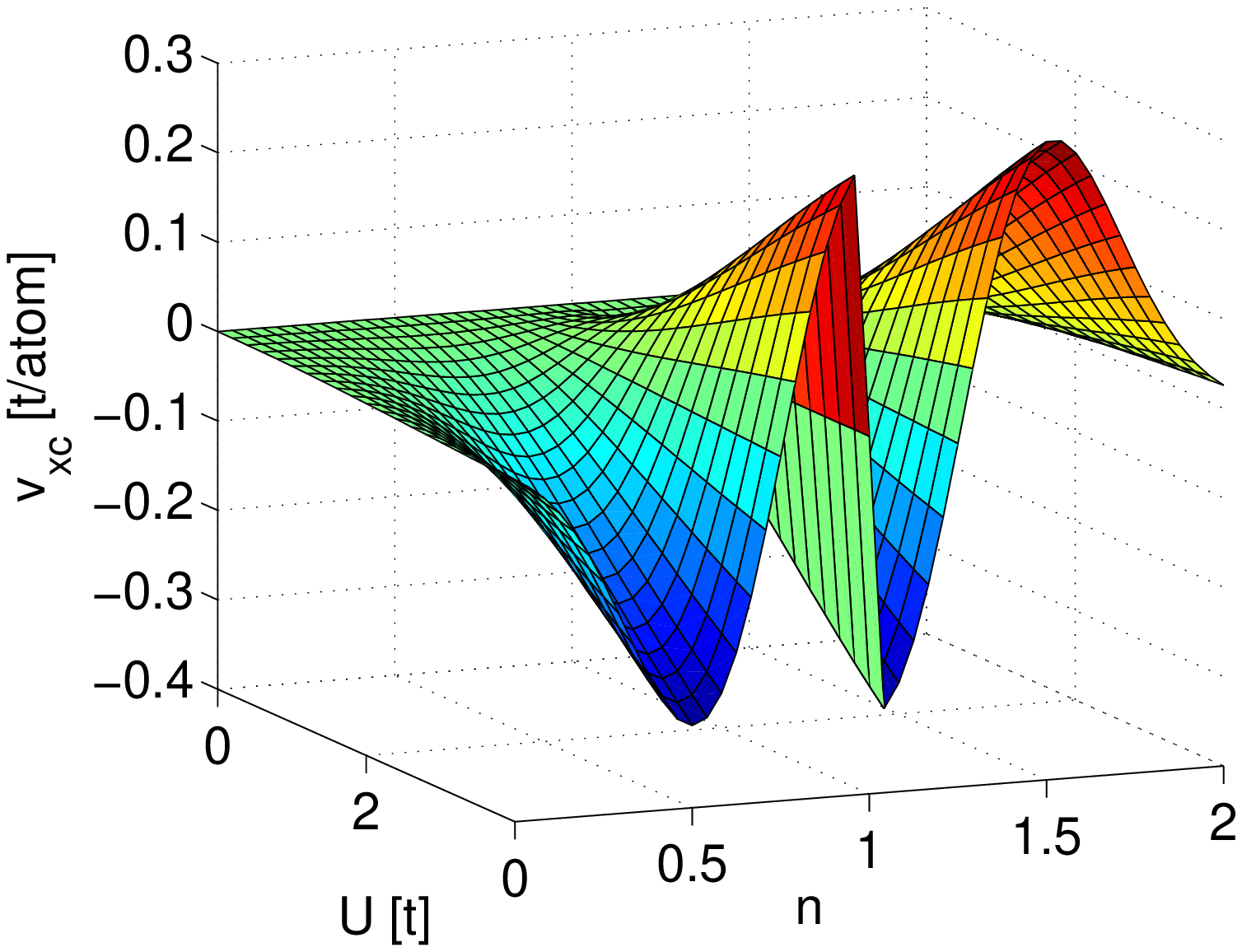}}
		\caption{\label{fig:srfcs} (Color online) The XC energy and potential fitted to the infinite graphene sheet with nearest-neighbour hopping for $S_z = 0$. Top row: 1NN hopping, bottom row: 3NN hopping (a/e) and (c/g): numerical XC energy and potential from ED (b/f) and (d/h): fitted analytic functions of form Eq. (\ref{eq:xcform}).}
\end{figure*}

The exchange-correlation energy and functional contain the many-body effects of the original problem. In the spirit of the XC functional for the 1D Hubbard chain by Lima \emph{et al.},\cite{Lima-Silva, Capelle} we choose to determine the functional within the local density approximation, first introduced by Sch\"{o}nhammer \emph{et al.} in Ref.~\onlinecite{Schonhammer-Gunnarsson-Noack}. The XC energy and potential on a given site thus depend only on the local density at that site and not, for example, on the density difference between the site and its neighbours. The total XC energy of the system is then simply obtained by summing over the contributions on each of the sites, 
\begin{equation} \label{eq:exctot} E_{\mathrm{xc}}({n_i}, U) = \sum_i e_{\mathrm{xc}}(n_i,U), \end{equation}
where $e_{\mathrm{xc}}(n_i,U)$ is the XC energy for a homogeneous reference system, defined as\\
\begin{equation} \label{eq:exc} e_{\mathrm{xc}}(n,U) = \frac{E(n,U)-E^{HF}(n,U)}{N_s}. \end{equation}
Here, $N_s$ is the number of sites in the reference system, $E(n,U)$ is the exact reference energy given in our case by the exact diagonalization of the Hubbard Hamiltonian at filling $n = (N_{\uparrow} + N_{\downarrow})/N_s$, where $N_{\sigma}$ is the number of electrons of the spin species $\sigma$, and $E^{HF}(n,U)$ is the energy of the same system calculated using the Hartree-Fock Hamiltonian, Eq.~(\ref{eq:HF}). For each number of electrons $N_{\uparrow} + N_{\downarrow}$, multiple values for the $z$-projection of the spin, $S_z = \frac{1}{2}(N_{\uparrow} - N_{\downarrow})$, were considered. More details on the exact diagonalization technique can be found in Refs.~\onlinecite{Lin} and \onlinecite{Lin-Gubernatis}.  

As the Hubbard model is numerically exactly diagonalizable only up to approximately 16-sites on present computers in the general nonsymmetric case, we chose a 12-atom supercell of the infinite graphene sheet as our reference system, shown in Fig.~\ref{fig:refsys}. As graphene flakes were thought to be the first system to apply the method to, no $k$-dependence was taken into account in the functional and the calculations were performed at the $\Gamma$ point with $\vec{k} = 0$, introducing no phase shift to the wave function when crossing the supercell boundary. 

The Hubbard model was solved for all combinations of $N_{\uparrow}$ and $N_{\downarrow}$ for $U = 0-4t$ both by exactly diagonalizing the many-body Hubbard Hamiltonian in the Fock basis, and in the single-electron framework using the the Hartree-Fock Hamiltonian.  The range of the $U$ values was chosen to be moderate but to well extend over the range $\approx t$ relevant for graphene calculations.\cite{Hancock} Fig.~\ref{fig:srfcs} shows the numerical data and the fit for the $S_z = 0$ case both for the XC energy and potential, on the top row for nearest-neighbour hopping and on the bottom row for up to third-nearest neighbour hopping. The XC energy was fitted to a function of form
\begin{equation} \label{eq:xcform} e_{\mathrm{xc}}(n,U) = \alpha_1(e^{-\alpha_2U^2}-1)e^{-(\alpha_3|n-1|-\alpha_4)^2}. \end{equation}
The values for the coefficients $\alpha_i$ are given in Table \ref{tab:coeff}, both in the case of nearest-neighbour and up to third nearest-neighbour hopping. The XC potential was then obtained as\\
\begin{equation} \label{eq:vxc} v_{\mathrm{xc}} = \frac{\partial e_{\mathrm{xc}}}{\partial n}. \end{equation}
To be rigorous, a functional derivative should be taken from $e_{\mathrm{xc}}$. In this case, however, this reduces to the usual derivative.  In general the fit is very good, although we note that the largest deviations occur for $U > 3t$ near and at half-filling, $n=1$. We also note that the derivative discontinuity of the XC energy, seen also in Fig.~\ref{fig:srfcs}cdgh, is correctly contained in our functional. 

The inclusion of the up to third-nearest neighbour (3NN) hopping changes the form of the XC energy surface only slightly (Fig.~\ref{fig:srfcs}, bottom row). The particle-hole symmetry of the nearest-neighbour (1NN) Hubbard model is lost through the inclusion of the further hoppings. This causes the XC energy surface to be no longer symmetric about $n = 1$, leading to a separate set of coefficients $\alpha_i$ for $n < 1$ and $n > 1$. The coefficients for $n > 1$ are given in Table~\ref{tab:coeff} as $\alpha_i^*$. Comparing the coefficients for the 1NN and 3NN cases, we see that the magnitude of the XC energy is slightly smaller for 3NN hopping as the coefficient $\alpha_1$ mainly determines this magnitude. Also, for 3NN hopping the values of the coefficients for $n < 1$ and $n > 1$ are very close to each other. 

The XC energy surfaces for higher values of $S_z = \frac{1}{2}, 1, \frac{3}{2}, \cdots$ were also calculated. As the number of accessible density values and thus reference data points decreases with increasing $S_z$ due to the finite supercell, the fitting of the XC energy surfaces become more and more ambiguous. In addition, in Section \ref{sec:res} we demonstrate that our approach captures well locally spin-polarized systems using only the $S_z = 0$ potential, although the parameters for the $S_z = \frac{1}{2}$ potential were also determined and are shown in Table~\ref{tab:coeff}. 

\begin{table}
\caption{\label{tab:coeff} The coefficients $\alpha_i$ for Eq. (\ref{eq:xcform}). 1NN stands for nearest-neighbour hopping and 3NN for up to third nearest-neighbour hopping. In the 3NN case the coefficients are given separately for $n < 1$ ($\alpha_i$) and $n >1$ ($\alpha_i^*$). Coefficients based on both $S_z = 0$ and $S_z = \frac{1}{2}$ reference data are given. See text for details. }
 \begin{tabular}{c|p{1.4cm}p{1.4cm}p{1.4cm}p{1.4cm}}
   & 1NN $S_z = 0$ & 1NN $S_z = \frac{1}{2}$ & 3NN $S_z = 0$ & 3NN $S_z = \frac{1}{2}$ \\
\hline
 $\alpha_1$ & 0.22554 &0.24732 &0.17094 & 0.15972 \\
 $\alpha_1^{*}$&- &- &0.17458 & 0.15071\\
$\alpha_2$ &0.05776 &0.04919 & 0.08322 &0.09122 \\
$\alpha_2^{*}$&- &- &0.08657 &0.10440 \\
$\alpha_3$&2.67381 &2.70556 & 2.77940 &2.98123 \\
$\alpha_3^{*}$&- &- &2.73657 & 2.82716 \\
$\alpha_4$&2.02445 &2.22949 & 2.09609&2.26728 \\
$\alpha_4^{*}$&- &- &2.02992 & 2.10459 \\
 \end{tabular}
 
\end{table}

\section{Results \label{sec:res}}

\begin{figure}
		\subfigure[]{\includegraphics[scale=0.37]{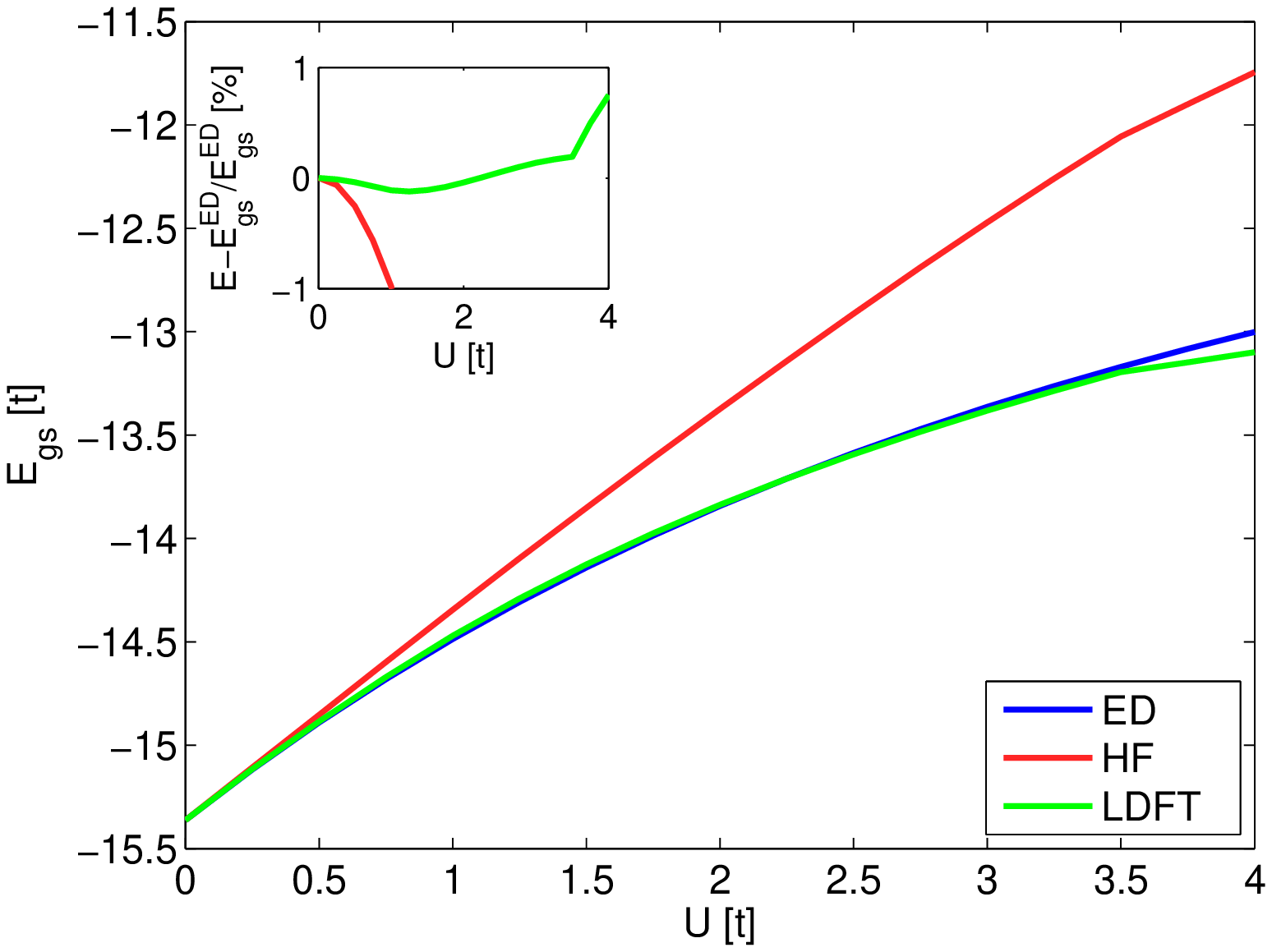}}
		\subfigure[]{\includegraphics[scale=0.37]{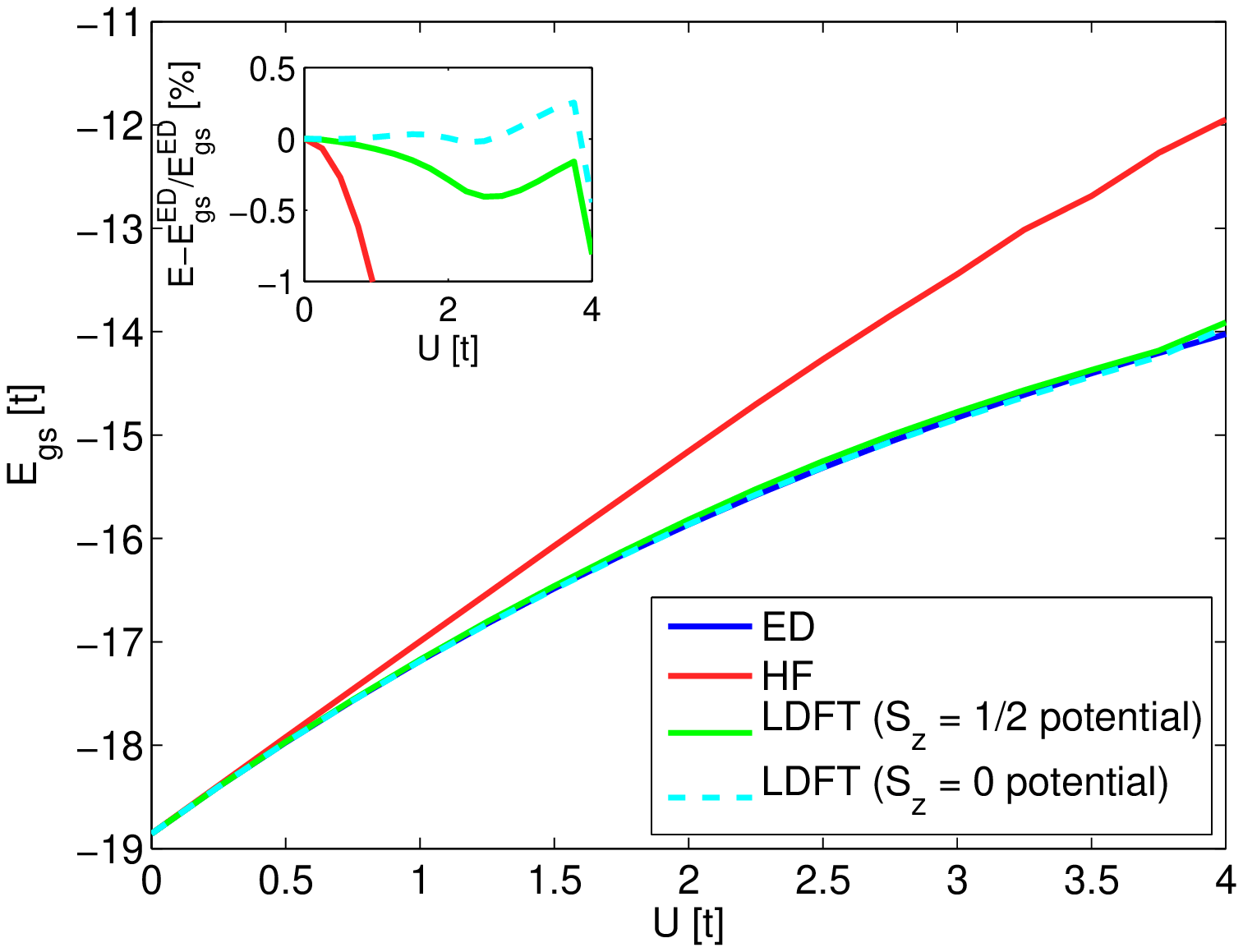}} 
		\subfigure[]{\includegraphics[scale=0.37]{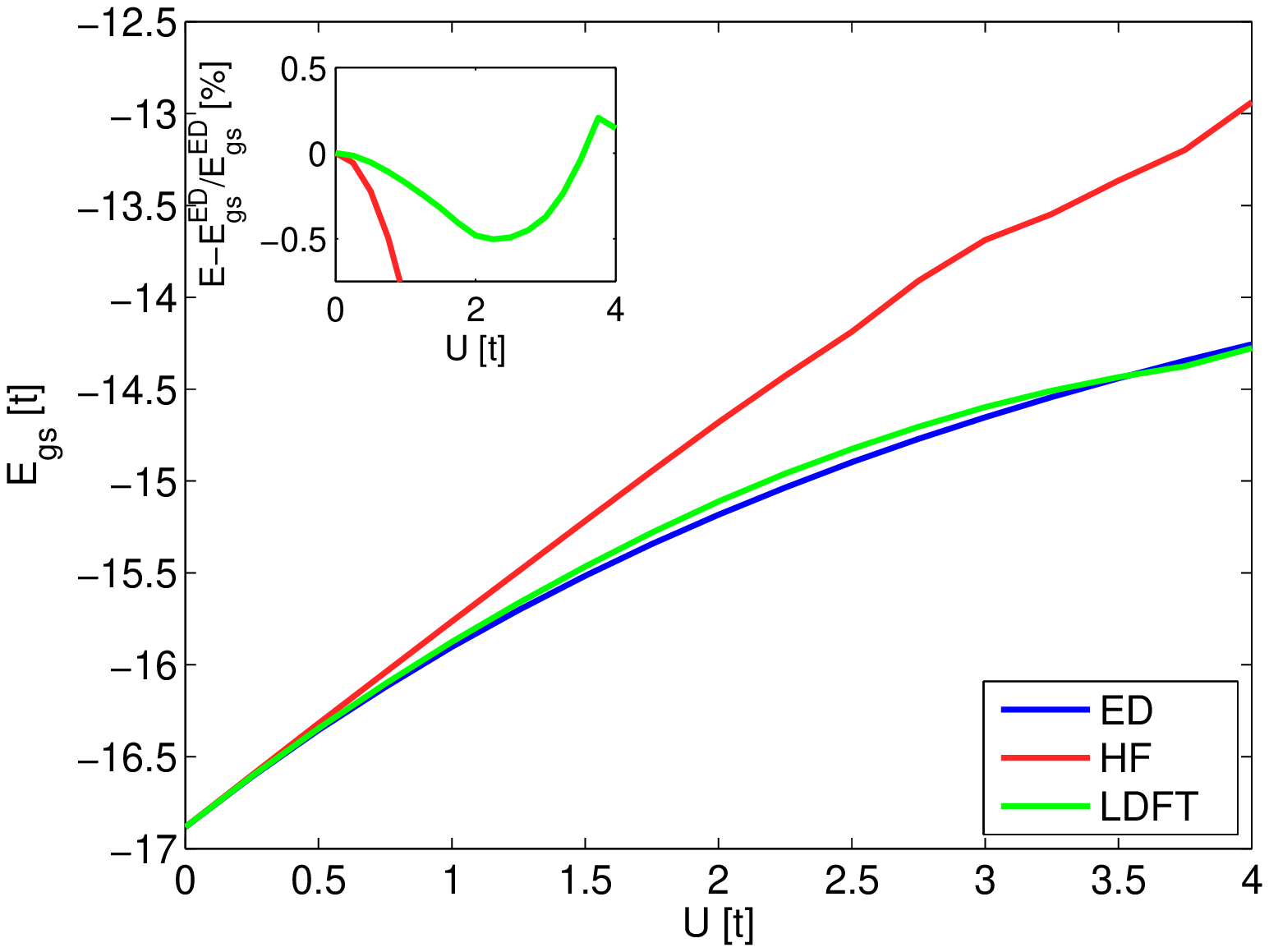}}
		\subfigure[]{\includegraphics[scale=0.37]{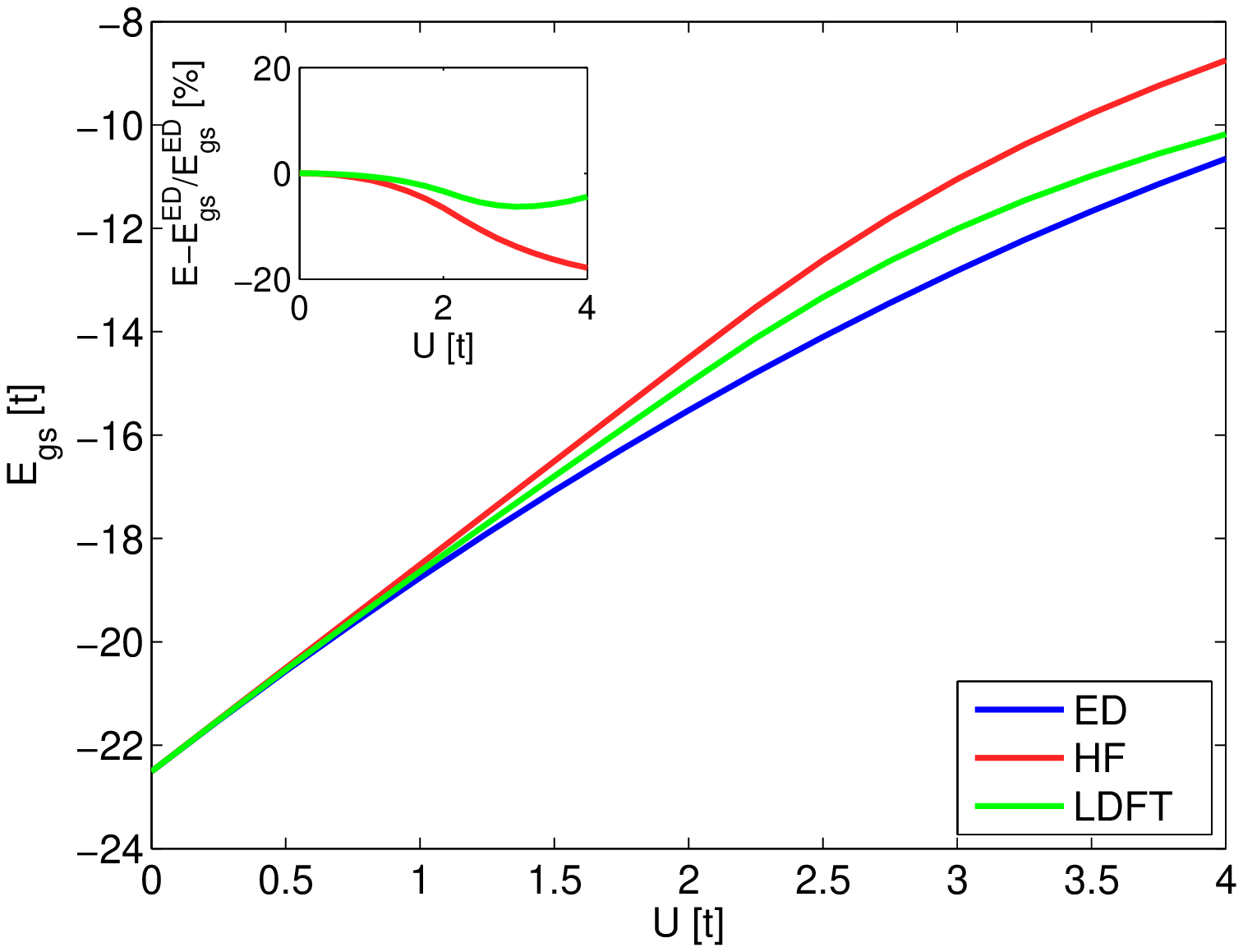}}
		\caption{\label{fig:C16egs} (Color online) A comparison between the ED, HF and LDFT ground-state energies of the C$_{16}$ flake (Fig.~(\ref{fig:refsys}). (a) At quarter-filling.  (b) With $S_z = \frac{1}{2}$ ($N_{\uparrow} = 6$, $N_{\downarrow} = 5$). LDFT calculation performed both using a potential fitted for $S_z = 0$ and $S_z = \frac{1}{2}$ reference systems. (c) At quarter-filling with an added on-site impurity potential $\epsilon = -|t|$ on the two middle sites of the structure (d) At half-filling. The insets show the relative errors (in percentage) with respect to the ED energy.}
\end{figure}

\subsection{Comparing LDFT to exact diagonalization}

As a first test for our functional, we compare the ground-state energy of a 16-atom graphene flake (Fig. \ref{fig:refsys}) calculated by exactly diagonalizing the Hubbard Hamiltonian, from the self-consistent solution of the Hartree-Fock Hamiltonian, Eq. (\ref{eq:HF}), andusing our LDFT method. We calculate the system at three different fillings: at quarter-filling ($N_{\uparrow} = N_{\downarrow} = 4$), slightly off quarter-filling ($N_{\uparrow} = 6$, $N_{\downarrow} = 5$) and at half-filling ($N_{\uparrow} = N_{\downarrow} = 8$), including only the nearest-neighbour hopping. Two different $S_z$ states, $S_z = 0$ and $S_z = \frac{1}{2}$ are thus included in this comparison. Additionally, we perturb the system by adding an on-site energy of magnitude $\epsilon = -|t|$ on the two middle sites of the structure. In general, introduction of the on-site energies gives to Eqs. (\ref{eq:Hub}) and (\ref{eq:HF}) a term of the form 
\begin{equation} \label{eq:onsite} \hat{V}_{\mathrm{ext}} = \sum_{k\sigma \in \{\mathrm{on-site}\}} \epsilon_k t_{k\sigma}^{\dagger}t_{k\sigma}, \end{equation}
where the sum runs over the site with an on-site energy and $\epsilon_k$ gives the magnitude of this energy. 

Fig.~\ref{fig:C16egs} shows the ground-state energy as a function of $U$ calculated using the three methods. The agreement between the LDFT and ED, apart from the half-filled case, is very good. This is also illustrated in the insets, which show the relative error in percentage with respect to the ED energy that is below 1 \% away from half-filling. At half-filling the agreement is worst, the relative error being $\approx$ 6 \%, at $U \approx 3 t$. This disagreement is partly due to the worse fit of the XC energy at half-filling. It is also worth noting that the difference between the two parameterizations of the potential for $S_z = 0$ and $S_z = \frac{1}{2}$ systems is actually in favour of using the $S_z = 0$ potential for the $S_z = \frac{1}{2}$ test system (see Fig.~\ref{fig:C16egs}b) but this is likely to be due to a coincidental cancellation of errors. On the other hand, this agreement is also to be expected as the $\hat{V}_{\mathrm{xc}}$ fits are very close to each other for these two spin states, see Table \ref{tab:coeff}. 

A comparison of the ground-state densities further resolves the issue of poorer results at half-filling. For the system without any on-site impurities, the exact density is uniform also for non-zero $U$. The Hartree-Fock solution, on the other hand, has an antiferromagnetic ground-state density with local spin polarizations. This characteristic of the Hartree-Fock solution is also contained in the LDFT density. In the half-filled case at $U = 3t$, the spin-resolved LDFT density ranges from 0.20 to 0.80 with the maximal local spin polarization $S_z^{i, \mathrm{max}} = 0.3$ for both spin up and spin down species. This separation of spin up and spin down densities is also seen in the quarter-filled case but with a smaller amplitude due to the lower average density. Thus, although the LDFT method corrects the ground-state energy quite accurately, the method fails to reproduce the homogenous ED density. This problem is due to the single-configuration wave functions in DFT.\cite{Harju}

\begin{figure}
		\subfigure[]{\includegraphics[scale=0.5]{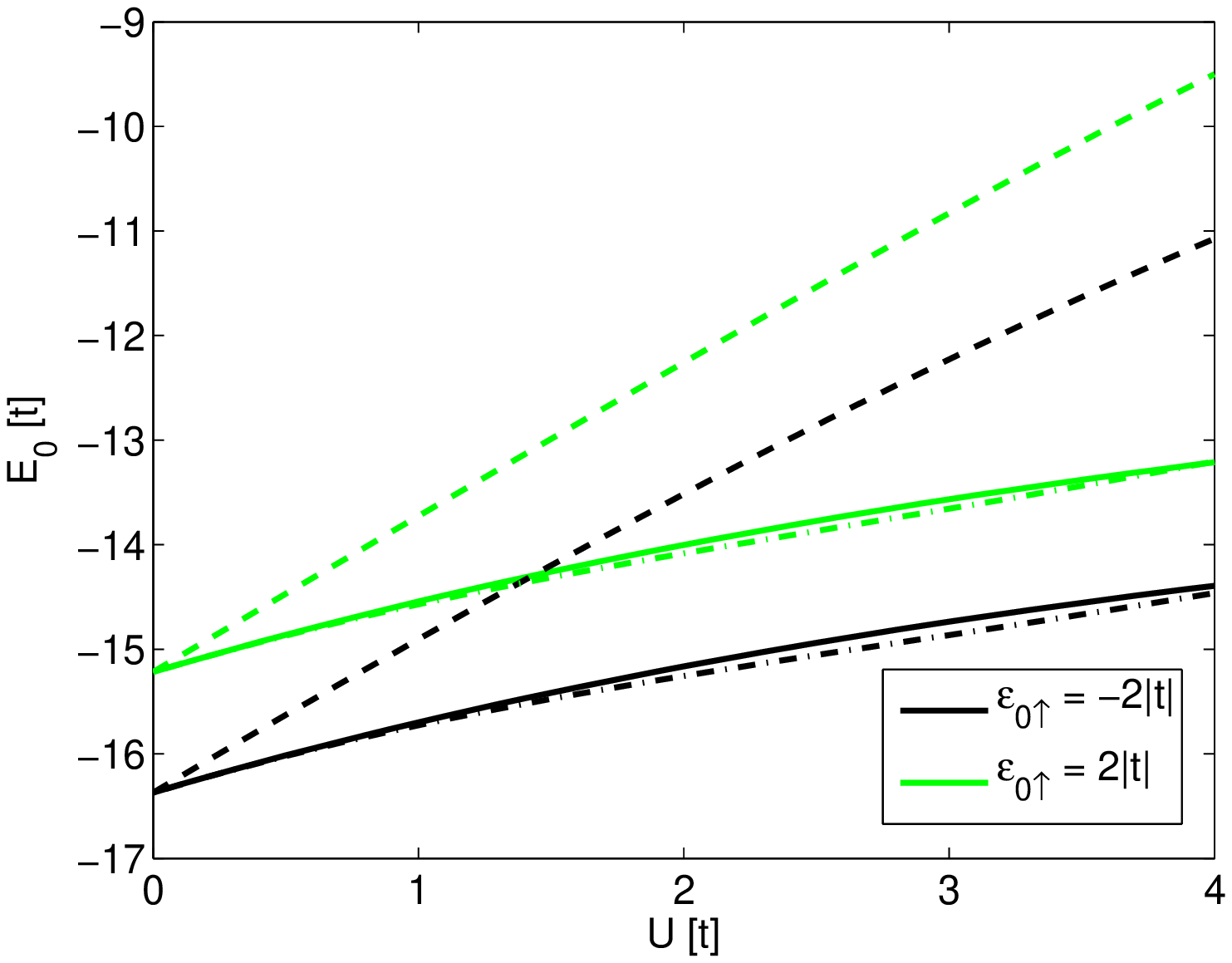}} 
		\subfigure[]{\includegraphics[scale=0.5]{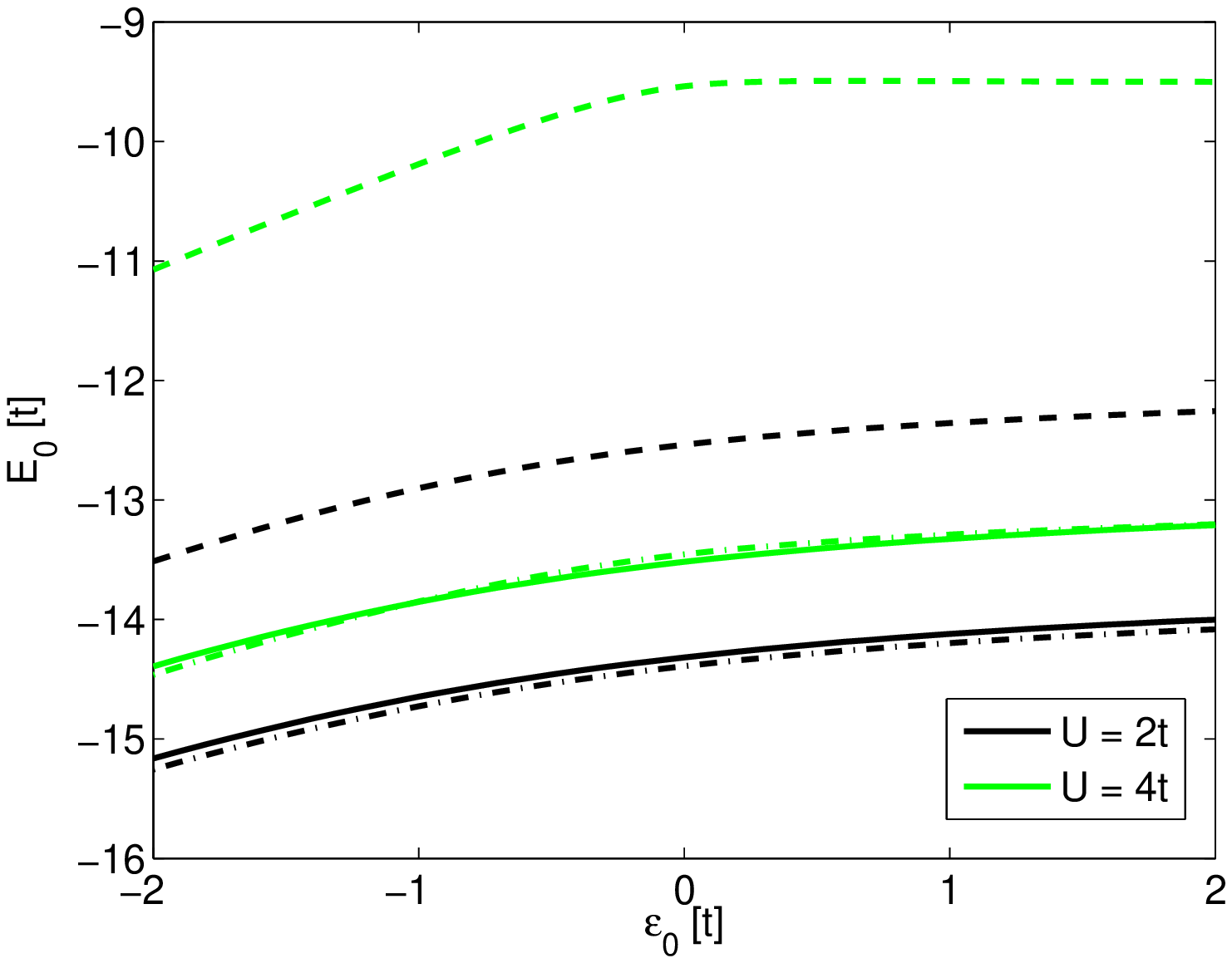}} 
		\caption{\label{fig:spine0} (Color online) The ground-state energy of an infinite graphene sheet (12-atom supercell) at quarter-filling, perturbed with a single spin-dependent on-site energy, using ED (full line), HF (dashed line) and third-nearest-neighbour LDFT (dash-dotted line). In (a) the strength of the on-site energy $\epsilon_0$ is varied at fixed $U$ and in (b) the on-site energy is fixed and $U$ varied.}
		\end{figure}

The effect of local spin polarizations can be studied by introducing a spin-dependent on-site energy as a perturbation to the model. This causes the ground-state density of the ED solution to become non-uniform. In the case of a quarter-filled infinite sheet with the 12-atom supercell (Fig.~\ref{fig:refsys}), $N_{\uparrow} = N_{\downarrow} = 3$, calculated including up to third nearest-neighbour hoppings with an on-site energy $\epsilon_0 = -2|t|$ applied on a single site for only the spin up species, the ED ground state exhibits spin-dependent occupations ranging from 0.14 to 0.60 at $U = 0$ and from 0.13 to 0.63 at $U = 3t$. As also the ED density is staggered, the LDFT density agrees better with it, the LDFT density varying at $U = 3t$ from 0.11 to 0.65. Fig.~\ref{fig:spine0}a shows the energy as a function of $U$ at two values of the on-site energy, and Fig.~\ref{fig:spine0}b the ground-state energy as a function of the on-site perturbation strength at two different non-zero values of $U$. The LDFT energies agree again very well with the exact solutions. 

To conclude this discussion,  we note that the computational effort of LDFT is roughly the same as HF. In comparison to ED, LDFT scales polynomially instead of exponentially with the system size.

\subsection{Triangular flakes}

Proceeding away from the simple, exactly solvable systems, we apply the LDFT method on triangular graphene nanoflakes. First we study  a still small triangle consisting of 22 sites that has  $S_z = 1$ in its ground-state at half-filling.\cite{Lieb, Potasz, Fajtlowicz} As a reference, we compare our LDFT calculation to a method we call ``partial diagonalization'' (PD) in which instead of diagonalizing the many-body Hamiltonian Eq.~(\ref{eq:Hub}) in the full Hilbert space, we only use the space spanned by the non-interacting ground-state and its up to double excitations. This method is thus equivalent to the singles-doubles configuration interaction method used in quantum chemistry. We calculate the non-interacting eigenstates of the system and occupy them according to the filling of the flake. We then singly and doubly excite the the system and calculate the elements of the Hamiltonian in the set of many-body states formed by the non-interacting ground-state and its excitations. This method is accurate at low values of $U$ up to $U\approx 2t$, ensured by comparing to ED calculations for 6- and 10-atom flakes. Flake sizes up to approximately 40 sites are easily accessible within this method. It should, however, be noted that the accuracy of PD at a fixed number of included excitations is expected to be reduced for larger systems as the correlation energy does not grow linearly with the system size.

\begin{figure}
		\subfigure[]{\includegraphics[scale=0.48]{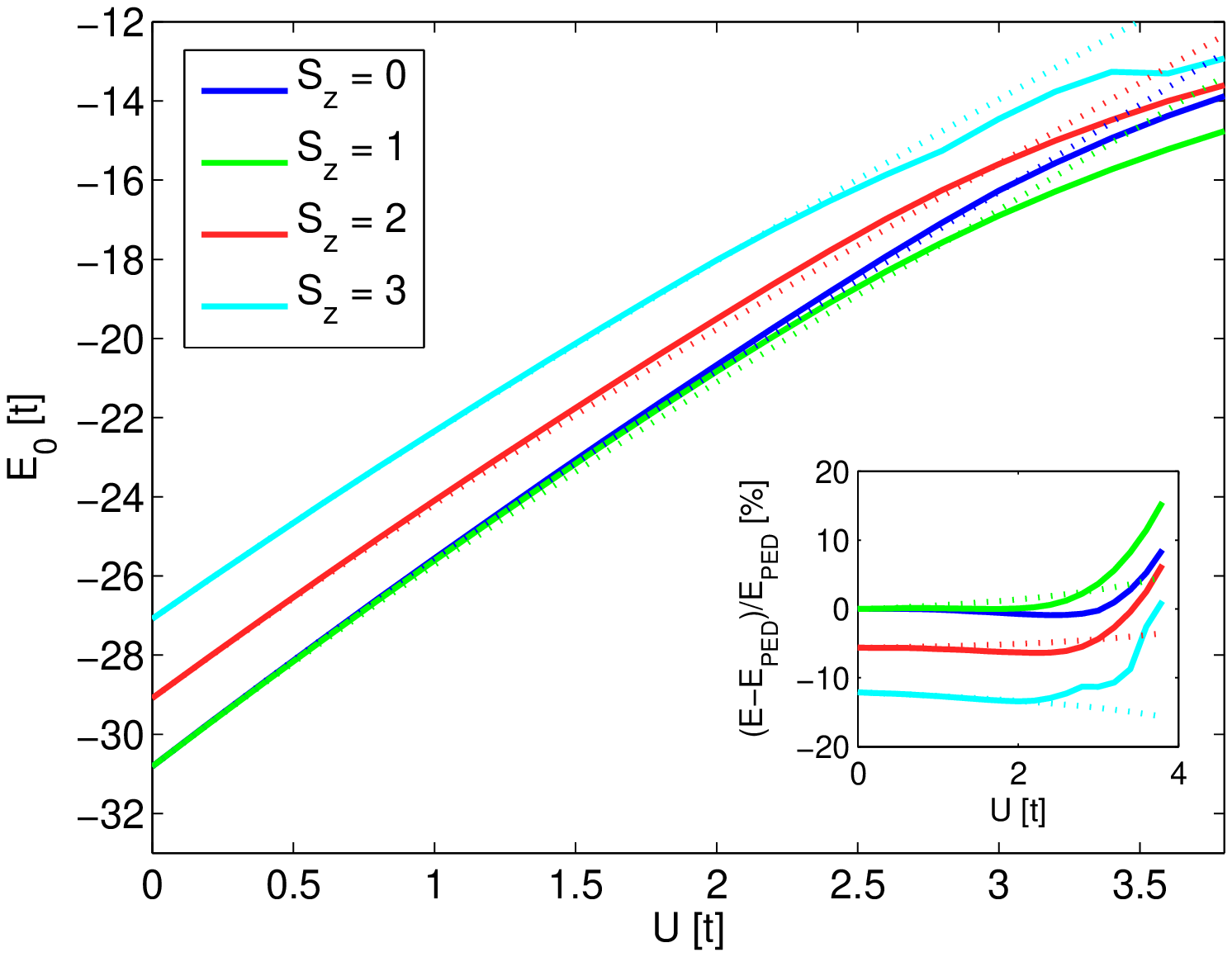}}
		\subfigure[]{\includegraphics[scale=0.25]{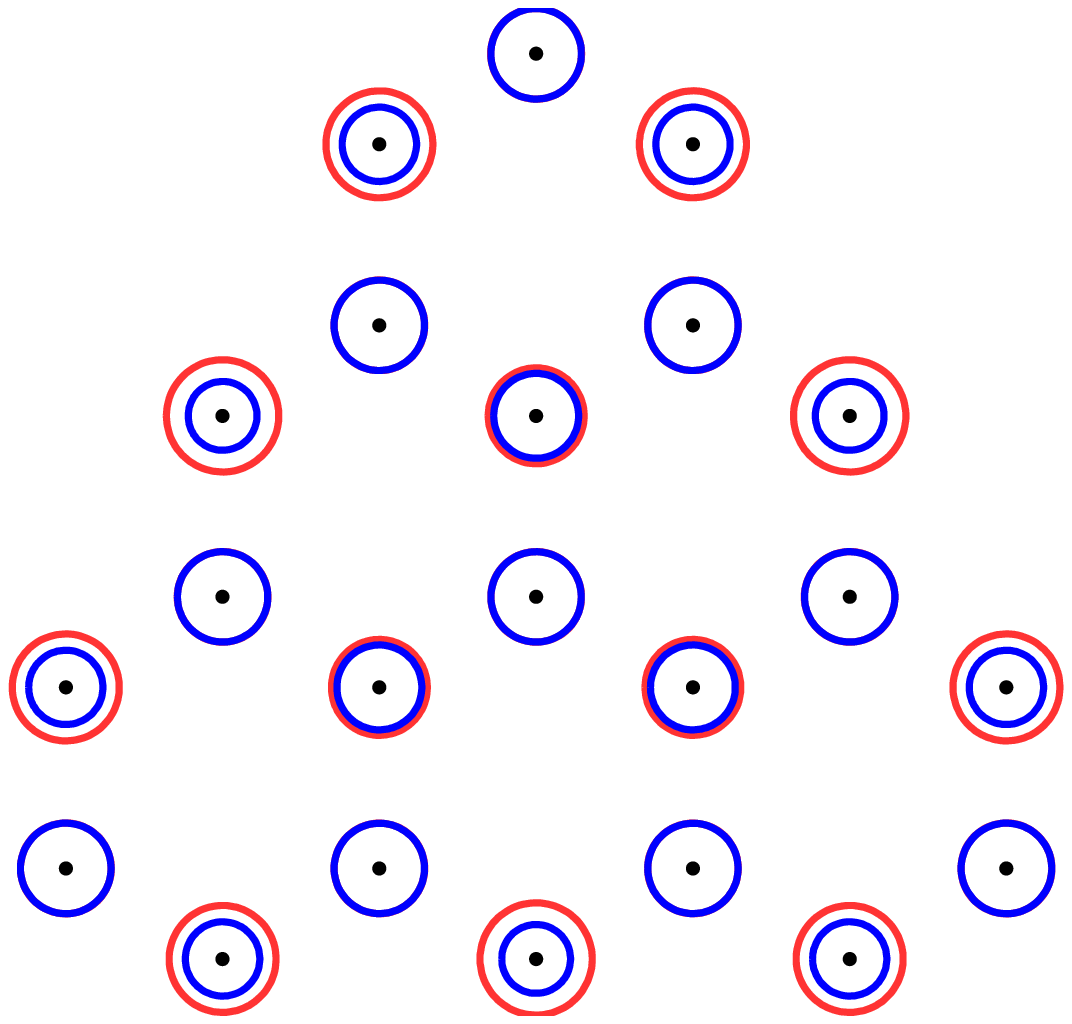}}
		\subfigure[]{\includegraphics[scale=0.25]{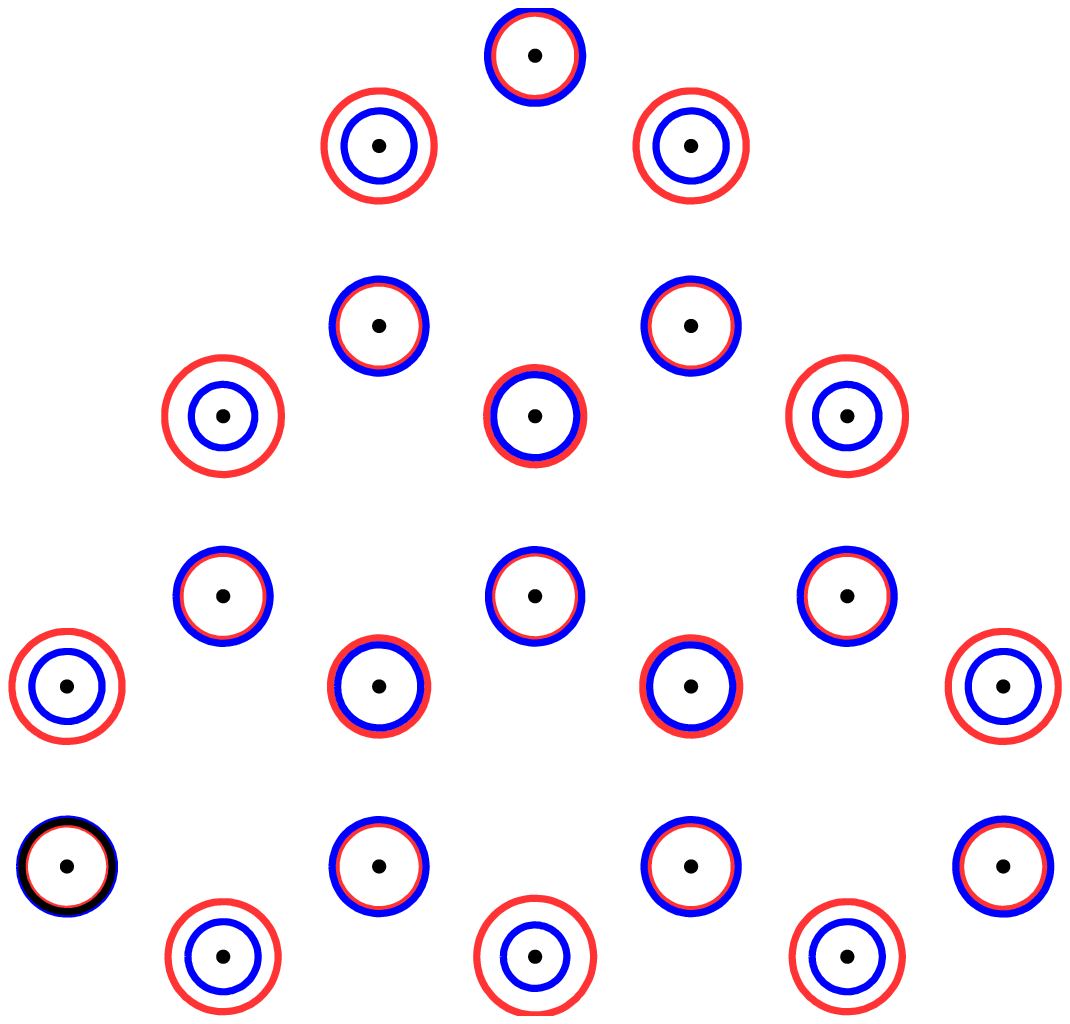}}
		\subfigure[]{\includegraphics[scale=0.25]{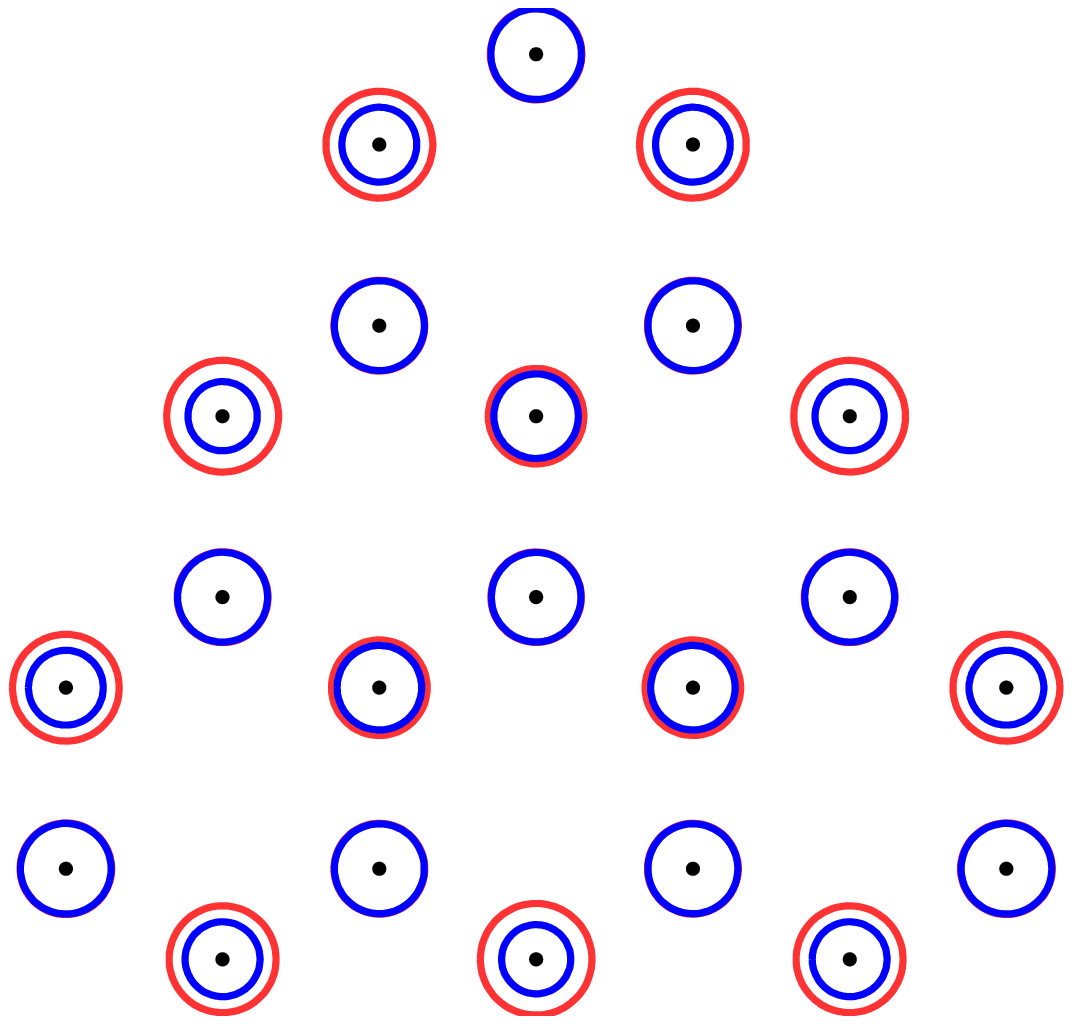}}
		\subfigure[]{\includegraphics[scale=0.25]{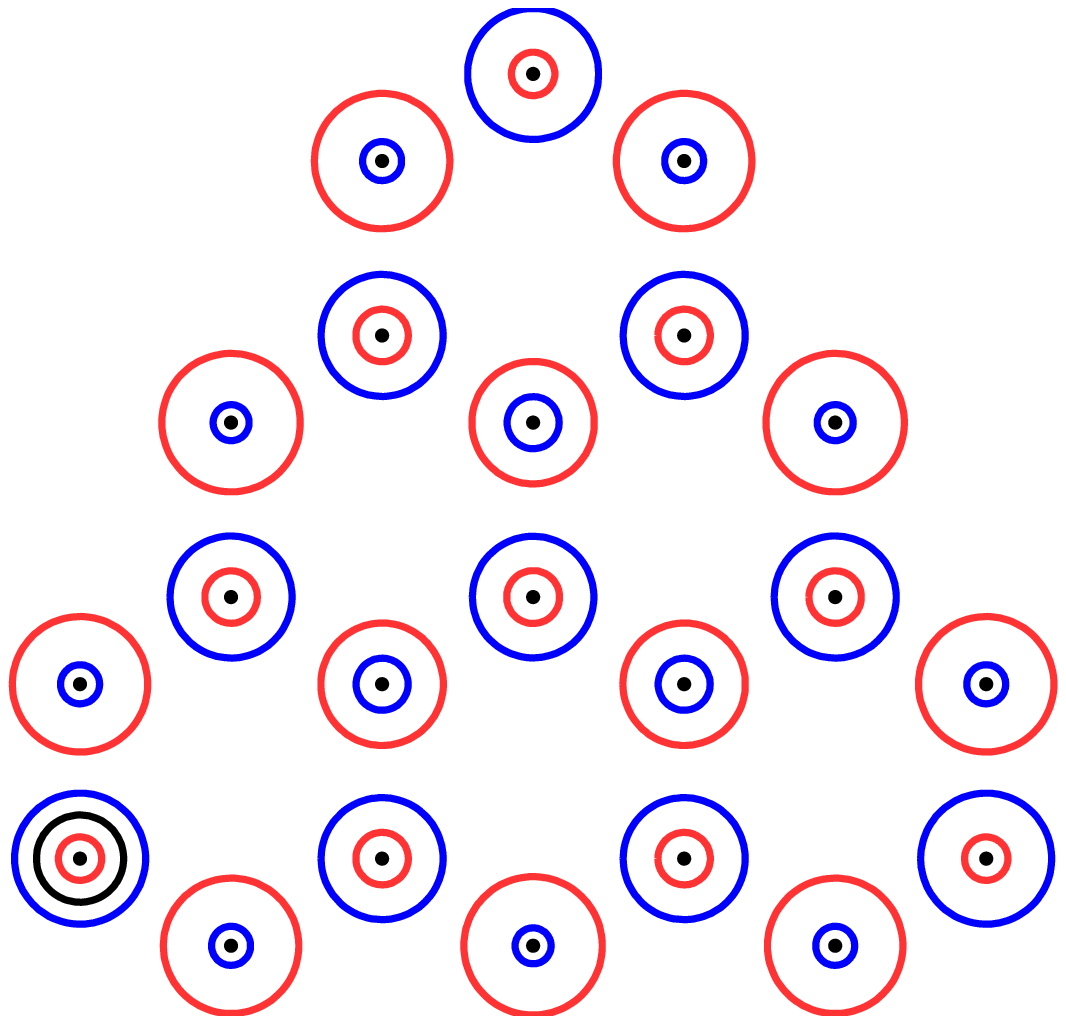}}
		\caption{\label{fig:c22} (Color online) Comparison of PD and LDFT within the nearest-neighbour hopping approach for the C$_{22}$ flake having a spin-polarized ground-state at half-filling. In (a) the comparison of the ground-state energy for different values of $S_z$. Solid line: LDFT, dotted line: PD. The inset shows the deviation in percentage from the PD $S_z = 0$ energy. In (b) and (c), the ground-state density at $U = t$ from PD and LDFT, respectively. In (d) and (e), the density at $U = 2.8t$. Red (light grey) and blue (dark) signify the up and down spin densities, respectively. Half-filling corresponds to the radius of the blue circles in the lower left corner in (b) and (d) and to the single black circle in (c) and (d), drawn to facilitate comparison. }
		\end{figure}

Fig.~\ref{fig:c22} compares the ground-state energy of the different $S_z$ states of the triangular C$_{22}$ flake calculated using PD and nearest-neighbour LDFT using only the $S_z = 0$ potential. In Fig.~\ref{fig:c22}a, we see that despite the use of the $S_z = 0$ potential, LDFT yields the correct $S_z$ value for the lowest-energy state at all values of $U$. The inset shows the deviation in percentage from the $S_z= 0$ state calculated using PD. At $U > 3t$ the PD energy becomes nearly linear in $U$ and we expect the LDFT energy to better describe the physics of the system, as for very large $U$ the ground-state should exclude doubly-occupied sites and the energy should saturate.\cite{Tasaki} This linearity in the PD solution at larger $U$ leads also to apparently large deviations in the inset of Fig.~\ref{fig:c22}a. Figs.~\ref{fig:c22}b and \ref{fig:c22}c show the ground-state density from PD and LDFT, respectively, for the lowest-energy state ($S_z = 1$) at $U = t$ and Figs.~\ref{fig:c22}d and \ref{fig:c22}e at $U = 2.8t$. The excess spin is localized on one of the sublattices and mostly on the outmost zigzag sites, a feature correctly captured by the LDFT solution and also reported in DFT calculations.\cite{Wang-Meng, Yazuev-Wang, Wang-Yazuev} For $U = t$, the deviation from the PD density is hardly noticeable, whereas for $U = 2.8t$ the deviation is clearly seen. Even at strong interaction, however, the qualitative features of sublattice-dependent polarization and spin localization on the outmost sites still remain.

In the 3NN triangular flakes, unlike the 1NN case,\cite{Lieb, Fajtlowicz, Potasz} there are no general predictions for the ground-state spin. As the hopping within the sublattice is relatively weak ($t_2 \approx 0.07 t$), it is not likely that the situation changes drastically from the system with only nearest-neighbour hopping. Third-nearest neighbour hopping again connects the two sublattices. According to a theorem by Lieb,\cite{Lieb} valid for Hubbard models on bipartite lattice with a repulsive interaction at half-filling, a 286-atom zigzag-edged triangle having 15 hexagons in the base row should have $S_z = 7$ in its ground-state.  The spin is given by the sublattice inbalance, $S_z = \frac{1}{2} | N_A - N_B|$, where $N_{A(B)}$ signifies the number of sites on sublattice $A$($B$). This is due to the fact that in the absence of interaction, the highest occupied single-electron state is 14-fold degenerate. Our LDFT approach indeed finds the state $S_z = 7$ to be of lowest energy for $U > 0.6t$. Due to the degeneracy, however, convergence problems are encountered at $U < 0.4t$ as the numerical solution is very unstable for the open shell iteration for lower values of $S_z$, and the lower-spin states appear to be energetically more favourable. 

\begin{figure}
 	\includegraphics[scale=0.5]{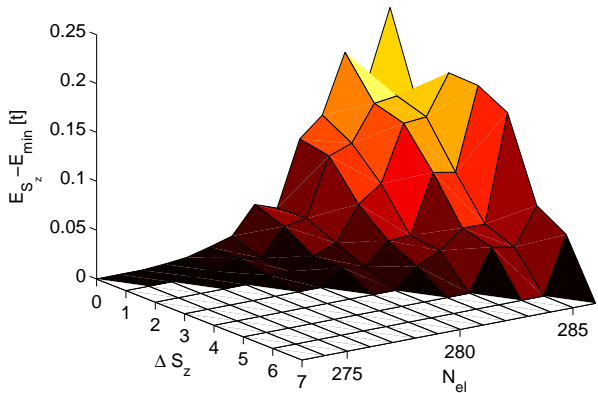}
	\caption{\label{fig:gap} The energy gap between the lowest-energy state  and the other $S_z$ states for the 286-atom triangular graphene flake at $U = t$ with different total numbers of electrons, $N_{\mathrm{el}} = N_{\uparrow} + N_{\downarrow}$. As the gap jumps when crossing the diagonal, the gap is shown only up to the $S_z$ value with minimal energy. The $x$-axis shows the difference in $S_z$ from the minimal spin, $S_{z,min} = 0$ for even $N_{\mathrm{el}}$ and $S_{z,min} = 1/2$ for uneven $N_{\mathrm{el}}$. } 
\end{figure}

In Fig.~\ref{fig:gap}, we show the energy gap at $U = t$ between the lowest-energy state of the 286-atom flake and the other $S_z$ values at and slightly below half-filling, the total number of electrons ranging from 273 to 286. The value of $U$ was chosen to be relevant to graphene calculations.\cite{Hancock, Hancock-Uppstu} At half-filling, the $S_z = 7$ has the lowest energy as expected from the Lieb theorem. Slightly off half-filling the energetically favourable $S_z$ decreases linearly with the total number of electrons but at the same time the magnitude of the gap decreases. Even though the states seem to lie very close to each other in energy, the order of magnitude of the gap at half-filling, 0.1$t \approx 0.27$ eV, transformed into temperature, $\approx 3100 K$, is high enough for us to predict that the high-spin ground-state should be observable in room temperature. Away from half-filling, on the other hand, the gap magnitude in temperature is of the order of room temperature, making the system an unlikely candidate for practical spintronics applications. 

An estimate for the accuracy of LDFT in determining the gap magnitude is obtained by considering the gap in C$_{22}$ (Fig.~\ref{fig:c22}). The gap between the $S_z = 1$ and $S_z = 0$ ground-states is 0.13 eV in PD and 0.06 eV in LDFT. LDFT thus seems to somewhat underestimate the gap but this only supports the conclusion on the rigidity of the ground-state against thermal fluctuations. As the C$_{22}$ is the smallest flake that has non-minimal spin in its ground-state, the accuracy of the LDFT predictions in these systems can not be compared to ED. To compare the gap in a diagonalizable system, the gap between the two lowest states ($S_z =  1/2$ and $S_z = 3/2$) in the smallest triangular flake, C$_{13}$, at half-filling was calculated using ED, PD, and LDFT at $U = t$ including up to third-nearest neighbour hopping. The gap magnitude in ED, 1.52 $t$, is thus an order of magnitude larger than in the spin-polarized flakes, and both PED and LDFT overestimate it but only by 3.3\% and 4.2\%, respectively.

\begin{figure}
 \includegraphics[scale=0.48,angle=90]{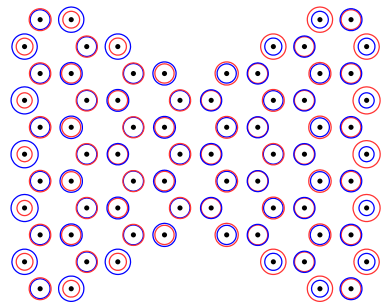}
 \caption{\label{fig:bowtie} (Color online) The ground-state density ($S_z = 0$) of a 78-atom bowtie flake  at half-filling, previously studied by Wang \emph{et al.} in Ref.~\onlinecite{Wang-Yazuev}. Up to third nearest-neighbour hopping is included and the value of $U$ is $t$. Red (light grey) and blue (dark) correspond to spin up and spin down densities, respectively.}  
\end{figure}

In order to show that the LDFT method is able to predict the spin densities also for more complex geometries, Fig.~\ref{fig:bowtie} shows the ground-state density of a bowtie-shaped flake obtained by combining two triangular flakes. Hopping up to third-nearest neighbours is included and the flake is calculated at half-filling and $U = t$. Unlike in the triangular flakes, the sublattice imbalance is zero in these structures and thus the ground-state has $S_z = 0$. Nevertheless, the spin structure is nontrivial as the spin up and spin down densities concentrate on the opposite sides of the structure. These structures have been previously proposed to function as spin logic devices by Wang \emph{et al.}, \cite{Wang-Yazuev} and a comparison with their density-functional calculation shows that LDFT indeed captures this non-trivial spin order.

Based on the results of this section, we conclude that the LDFT approach captures also the essential features of systems with a non-zero total spin in the ground-state or with large local spin polarizations.

\section{Conclusions \label{sec:concl}}

Lattice-density functional theory is a promising candidate to be used instead of the mean-field Hartree-Fock approximation for graphene systems too large for exact methods. The exact ground-state energies are accurately reproduced, and  correct spin properties for large graphene flakes are found. Local density approximation (LDA) is enough to capture these spin properties and no need for an explicitly spin-dependent potential was found. The ground-state densities are antiferromagnetic with local spin-polarization like the HF solution and the computational effort is of the same order as in the mean-field Hartree-Fock approximation.  

The method could be applied to transport calculations, for instance for graphene nanoribbons. Also, the method could be extended into time domain in the form of an adiabatic time-dependent lattice density-functional theory and used to study dynamic phenomena such as the possibility of spin-charge separation above one dimension.

\section*{Acknowledgments}
We thank Prof. Risto Nieminen for useful discussions and gratefully acknowledge the support of Academy of Finland through its Centers of Excellence Program (2006-2011). This work was also partially supported by the Aalto-Nokia collaboration. In addition, M. I. acknowledges the support from the Finnish Doctoral Programme in Computational Sciences FICS.


%
\end{document}